\documentclass[aps,prb,floatfix,epsfig,twocolumn,showpacs,preprintnumbers]{revtex4}
\UseRawInputEncoding
\usepackage{amsfonts}
\usepackage{amssymb}
\usepackage{graphicx,axodraw2}
\usepackage{amsmath}
\usepackage{tablefootnote}
\usepackage{setspace}
\begin{document}

\title{Electronic structure of van der Waals ferromagnet CrI$_{3}$ from self consistent vertex corrected GW approaches}

\author{Andrey L. Kutepov\footnote{e-mail: akutepov@bnl.gov}}

\affiliation{Condensed Matter Physics and Materials Science Department, Brookhaven National Laboratory, Upton, NY 11973}

\begin{abstract}
Electronic structure of layered van der Waals ferromagnet CrI$_{3}$ is studied with self consistent diagrammatic approaches beyond GW approximation. Considerable improvement in the calculated band gap as compared to the non-self-consistent G0W0 results has been found. Certain spectral features in the valence bands discovered recently by the angle resolved photoemission spectroscopy, are reproduced better when we use full frequency dependent self energy. Density functional theory and quasiparticle self-consistent GW method which are based on frequency-independent self energy are unable to resolve these features. Non-locality effects in the diagrams beyond GW approximation are large for both polarizability and self energy. This finding can potentially have an impact on the development of methods like GW+DMFT.
\end{abstract}

\maketitle


\section*{Introduction}
\label{intr}

Magnetic van der Waals material CrI$_{3}$ represents considerable interest in view of its promising applications in spintronics. It possesses some remarkable properties which include, for instance, the preservation of magnetic order down to a single layer.\cite{nature_546_270,scirep_9_13599} The bi-layer of this material shows anti-ferromagnetic ordering whereas its mono-layer, three-layer and bulk are all ordered ferromagnetically.\cite{nature_546_270} It is important to understand these (and other) properties from the theoretical point of view in order to be able to explain already known properties or even to predict new ones in this class of materials. The key to understand them is their electronic structure.

Electronic structure of CrI$_{3}$ was studied both experimentally\cite{scirep_10_15602,jap_36_1259} and theoretically.\cite{prm_3_031001,jmcc_3_12457,njp_21_053012,nanolett_18_3844,ncomm_10_2371,prb_101_241409} As it seems, there is a general consensus that basic features of it (such as band gap) are similar in bulk material and in thin film.\cite{scirep_10_15602,jmcc_3_12457} However, there is still no consensus on the reasons of apparent inconsistency between experimental and theoretical values of the band gap in CrI$_{3}$.

In the bulk CrI$_{3}$, optical measurement\cite{jap_36_1259} resulted in the optical gap of 1.24 eV. Recent ARPES (angle resolved photoemission spectroscopy) measurements\cite{scirep_10_15602} reported the electronic band gap of about 1.3 eV. Normally, one would think that optical gap should be a bit smaller than electronic because of the excitonic effects. Therefore, the above two values are consistent if we assume that the exciton binding energies are on the scale of 0.1 eV. In theory, there are issues on the larger scale. In density functional theory (DFT) calculations, the band gap is 0.78 eV.\cite{prb_101_241409} This value corresponds exactly to what one would expect from DFT: underestimation of the gap by 30-50\%. The problem reveals itself when we try to improve DFT band gap. Routinely, it is done by applying the so called one-shot (non self consistent) GW approximation (G0W0). In a vast majority of semiconductors, G0W0 improves the DFT band gap considerably\cite{prb_93_115203} with remaining small underestimation up to 10-15\%. However, when applied to the mono-layer of CrI$_{3}$, G0W0 results in the band gap 2.59--2.76 eV.\cite{ncomm_10_2371,jmcc_8_8856} It is important to note that reported G0W0 calculations of CrI$_{3}$ monolayer used DFT+U as a starting point. If we assume that bulk and mono-layer band gaps of CrI$_{3}$ are not very different, the reported G0W0 results for the mono-layer exceed considerably the experimental value. Which, most likely, is the case because authors of both works also reported very strong excitonic effects with exciton binding energies up to 1.5 eV. Formally, the presence of strong excitons could explain the value of the optical gap but it doesn't explain the value of the electronic gap. Nor, does it explain the small difference between optical and electronic gaps in experiments. However, it suggests that the electronic gap obtained in G0W0 calculations should be a subject of a strong renormalization if one includes diagrams beyond GW approximation in the evaluation of the electronic gap. For instance, if one uses Bethe-Salpeter equation (BSE) instead of random phase approximation (RPA) in the evaluation of polarizability and then applies the corresponding screened interaction W in the evaluation of the GW diagram, G0W0 band gap might be much smaller. Thus, the results obtained in Refs. [\onlinecite{ncomm_10_2371,jmcc_8_8856}] suggest to study the effect of higher order diagrams (vertex corrections) on the electronic structure of CrI$_{3}$.

An important step forward in elucidating the electronic structure of CrI$_{3}$ (and related materials) was done by Lee et al.\cite{prb_101_241409} In their work, the hybrid method QSGW80\cite{jjap_55_051201} was used. The QSGW80 approach consists in empirical mixing of QSGW (quasiparticle self-consistent GW) self energy and LDA (local density approximation) exchange-correlation potential: $\Sigma_{QSGW80}=0.8\Sigma_{QSGW}+0.2V^{xc}_{LDA}$. As authors of Ref. [\onlinecite{prb_101_241409}] argue, the mixing effectively corrects the underestimation of screening in QSGW method. Formally, the QSGW80 approach should be considered as a semiempirical one but it allows to improve the calculated electronic structure of simple semiconductors considerably.\cite{jjap_55_051201,prm_2_013807} For CrI$_{3}$, application of QSGW80 without spin-orbit coupling (SOC) resulted in the band gap 2.23 eV,\cite{prb_101_241409} whereas calculations with perturbative (after the self-consistency was reached) inclusion of SOC resulted in the band gap 1.68 eV. Thus, SOC renormalization of the electronic structure of CrI$_{3}$ is noticeable. Unfortunately, authors of Ref. [\onlinecite{prb_101_241409}] do not report the gap value obtained with standard QSGW, i.e. without admixture of LDA exchange-correlation potential. So, it is hard to say about the actual effect of it. QSGW80 is constructed in such a way that it empirically enhances the screening which is underestimated by QSGW. So, the mere fact that Lee et al. use QSGW80 instead of QSGW suggests an importance of higher order diagrams which would directly (instead of empirically) address the issue of insufficient screening in QSGW.

Authors of Ref. [\onlinecite{prb_101_241409}] also make an interesting research into the importance of nonlocality of self energy. Namely, by direct comparison of DFT+U and QSGW80 calculations they observe that DFT+U approach cannot mimic the QSGW80 results because of single-site approximation inherent to DFT+U. Obviously, this analysis of nonlocality of self energy in CrI$_{3}$ (and related materials) makes direct impact on the validity of other methods based on the single site approximation (like DFT plus dynamical mean field theory (DMFT)) when applied to this class of materials.

Motivated by the above cited works, this study focuses on application of the diagrammatic approaches which go beyond GW approximation, i.e. directly (and self-consistently) include vertex corrections. In this way, we estimate step by step the effect of the first order vertex correction and then the effect of replacing the first order diagram for polarizability by solving BSE for it. We also apply QSGW and, by doing this, we answer the question (though using different codes) on the difference between QSGW and QSGW80. Also, the effect of the SOC is studied directly. Namely, fully relativistic (FR) approach (Dirac's equation based) is used along with the scalar-relativistic (SR) approach in order to estimate SOC effect directly and compare it with the perturbative estimate made in Ref. [\onlinecite{prb_101_241409}]. We extend the study of non-local effects conducted by Lee et al. in [\onlinecite{prb_101_241409}] by investigating non-local contribution of the diagrams beyond GW. It is done by directly evaluating them using a full setup (all functions are \textbf{k}-dependent) and a simplified setup where we assume the local (single site) approximation. Our study, therefore, has an explicit impact on the development of the methods like GW+DMFT\cite{prl_90_086402,prb_94_201106,prm_1_043803,prr_2_013191,prx_10_041047,npj_1_16001,cpc_244_277} where one assumes the single site approximation for the DMFT part.

The paper begins with a brief discussion of the distinctive features of the methods used in this work and the setup parameters for the calculations (the first section). The second section provides principal
results obtained for the electronic structure of CrI$_{3}$. The third section presents the results of the investigation into the importance of non-local effects for higher order diagrams. The conclusions are given afterwards.

\section*{Methods and calculation setups}\label{meth}

\begin{figure}[t]
\begin{center}\begin{axopicture}(200,56)(0,0)
\SetPFont{Arial-bold}{28}
\SetWidth{0.8}
\Text(10,10)[l]{$\Psi$ =}
\Text(35,10)[l]{$-\frac{1}{2}$}
\GCirc(75,10){20}{1}
\Photon(55,10)(95,10){2}{5.5}
\Text(110,10)[l]{+}
\Text(128,10)[l]{$\frac{1}{4}$}
\GCirc(160,10){20}{1}
\Photon(160,-10)(160,30){2}{5.5}
\Photon(140,10)(180,10){2}{5.5}
\end{axopicture}
\end{center}
\caption{Diagrammatic representation of $\Psi$-functional which includes the simplest non-trivial vertex. First diagram on the right hand side stands for scGW approximation, whereas total expression corresponds to sc(GW+G3W2) approximation.}
\label{diag_Psi}
\end{figure}

\begin{figure}[t]
\begin{center}\begin{axopicture}(200,56)(0,0)
\SetPFont{Arial-bold}{28}
\SetWidth{0.8}
\Text(10,10)[l]{$P$  =}
\Photon(48,10)(55,10){2}{2.5}
\GCirc(75,10){20}{1}
\Photon(95,10)(102,10){2}{2.5}
\Text(120,10)[l]{$-$}
\Photon(143,10)(150,10){2}{2.5}
\GCirc(170,10){20}{1}
\Photon(190,10)(197,10){2}{2.5}
\Photon(170,-10)(170,30){2}{5.5}
\end{axopicture}
\end{center}
\caption{Diagrammatic representation of irreducible polarizability in the simplest vertex corrected scheme sc(GW+G3W2).}
\label{diag_P}
\end{figure}

\begin{figure}[t]
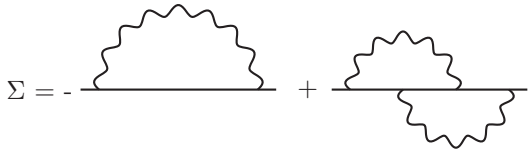

\begin{center}\begin{axopicture}(200,56)(0,0)
\SetPFont{Arial-bold}{28}
\SetWidth{0.8}
\Text(10,10)[l]{$\Sigma$  = -}
\Line(38,10)(112,10)
\PhotonArc(75,10)(30,0,180){2}{8.5}
\Text(120,10)[l]{$+$}
\Line(133,10)(207,10)
\PhotonArc(160,10)(20,0,180){2}{6.5}
\PhotonArc(180,10)(20,180,360){2}{6.5}
\end{axopicture}
\end{center}
\caption{Diagrammatic representation of self energy in the simplest vertex corrected scheme sc(GW+G3W2).}
\label{diag_S}
\end{figure}

All calculations in this work were performed using code FlapwMBPT.\cite{flapwmbpt_1} Recently, a few updates were implemented in the code.\cite{prb_103_165101,jcm_33_235503} For DFT calculations, we used the local density approximation (LDA) as parametrized by Perdew and Wang.\cite{prb_45_13244} In this study we use scGW method and two self-consistent vertex corrected schemes (see below). They are based on the L. Hedin's theory.\cite{pr_139_A796} ScGW and one of the vertex corrected schemes, sc(GW+G3W2)\cite{arx_2105_03770}, can also be defined using $\Psi$-functional formalism of Almbladh et al.\cite{ijmpb_13_535} Corresponding $\Psi$-functional which includes vertex corrections is shown in Fig. \ref{diag_Psi}. In Fig. \ref{diag_Psi}, the first diagram corresponds to GW approximation, whereas the sum of the first and the second diagram represents sc(GW+G3W2) approximation. Diagrammatic representations for irreducible polarizability (Fig. \ref{diag_P}) and for self energy (Fig. \ref{diag_S}) in scGW and in sc(GW+G3W2) follow from the chosen approximation for $\Psi$-functional.

\begin{figure}[t]
\begin{center}\begin{axopicture}(200,56)(0,0)
\SetPFont{Arial-bold}{28}
\SetWidth{0.8}
\Text(-18,10)[l]{$\Delta P$  =}
\Photon(13,10)(17,10){2}{1.5}
\GCirc(37,10){20}{1}
\Photon(37,-10)(37,30){2}{5.5}
\Photon(57,10)(61,10){2}{1.5}
\Text(66,10)[l]{$+$}
\Photon(78,10)(82,10){2}{1.5}
\GCirc(102,10){20}{1}
\Photon(95,-8)(95,28){2}{4.5}
\Photon(109,-8)(109,28){2}{4.5}
\Photon(122,10)(126,10){2}{1.5}
\Text(131,10)[l]{$+$}
\Photon(144,10)(148,10){2}{1.5}
\GCirc(168,10){20}{1}
\Photon(158,-6.5)(158,26.5){2}{4.5}
\Photon(168,-10)(168,30){2}{4.5}
\Photon(178,-6.5)(178,26.5){2}{4.5}
\Photon(188,10)(192,10){2}{1.5}
\Text(195,10)[l]{$+ ...$}
\end{axopicture}
\end{center}
\caption{Ladder sequence of diagrams for the vertex correction to polarizability in sc(BSE:P$@$GW+G3W2) approach.}
\label{bse}
\end{figure}

\begin{figure}[t]
\begin{center}       
\includegraphics[width=8.5 cm]{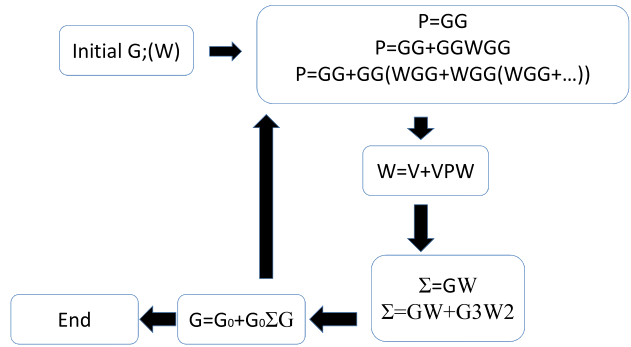}
\caption{Flowchart of scGW, sc(GW+G3W2), and sc(BSE:P$@$GW+G3W2) calculations. All equations are presented using symbolic notations. In the expressions for polarizability, first equation corresponds to scGW, second equation is used in sc(GW+G3W2), and the third one in sc(BSE:P$@$GW+G3W2). In the expressions for self energy, first equation corresponds to scGW, and the second one to both sc(GW+G3W2) and sc(BSE:P$@$GW+G3W2). $G_{0}$ stands for Green's function in Hartree approximation. Any calculation begins with self-consistent DFT iterations where the basis set is formed and the initial approach for G is generated. Iterations of scGW method use this initial Green's function as an input in order to start. During scGW iterations, G is updated and screened interaction W is generated. Both G and W serve as an input to start iterations of sc(GW+G3W2) or sc(BSE:P$@$GW+G3W2) approaches. sc(BSE:P$@$GW+G3W2), being computationally most demanding, can be run after a few iterations of sc(GW+G3W2), which can save computer time. In spin-polarized calculations, an external magnetic field is applied at the first iteration to create initial spin splitting.}
\label{flow}
\end{center}
\end{figure}

The second vertex corrected scheme which we use in this work is the scheme G according to the classification introduced in Ref. [\onlinecite{prb_94_155101}]. This scheme differs from sc(GW+G3W2) in the evaluation of polarizability: Bethe-Salpeter equation is used in the scheme G. In this case, the second term on the right hand side of Fig. \ref{diag_P} is replaced with an infinite sequence of diagrams (ladder diagrams) so that the vertex correction to polarizability can be represented as in Fig. \ref{bse}. Diagrammatic representation of self energy is the same in both vertex corrected schemes used in this work. For convenience, let us here introduce a new abbreviation for the scheme G: sc(BSE:P$@$GW+G3W2). In this abbreviation, the part after the symbol $@$ stands for diagrammatic representation of self energy, whereas the part before the symbol $@$ says that polarizability is evaluated from BSE. The rational of using sc(BSE:P$@$GW+G3W2) is to directly check the relative importance of excitonic effects on the evaluated electronic band structure. It is important to mention that our implementation\cite{prb_94_155101} of BSE uses full frequency dependence of screened interaction W opposite to a common approximation\cite{prl_80_4510,prb_78_085103} where one uses static (frequency independent and taken at zero frequency) screened interaction W. As one can deduce from its construction, scheme sc(BSE:P$@$GW+G3W2) is not $\Psi$-derivable (as opposed to scGW or sc(GW+G3W2)) and, therefore, is not conserving. However, evaluation of polarizability in sc(BSE:P$@$GW+G3W2) follows (at least approximately) its definition as being a functional derivative of electronic density with respect to full electrostatic potential, which is the foundation of BSE. Therefore, scheme sc(BSE:P$@$GW+G3W2) also has certain strong principle built in its construction. As it is evidenced in Ref. [\onlinecite{prb_95_195120}] it usually results in better band gaps as compared to sc(GW+G3W2). More details about properties of vertex corrected schemes one can find in Refs. [\onlinecite{prb_94_155101,prb_96_035108}].

Technical details of the GW part were described in Refs. [\onlinecite{prb_85_155129,cpc_219_407}]. Detailed account of the algorithms for sc(GW+G3W2), sc(BSE:P$@$GW+G3W2), and also for other vertex corrected schemes can be found in Refs. [\onlinecite{prb_94_155101,prb_95_195120,prb_96_035108,arx_2105_03770}]. Brief account of the implementation of BSE also is provided in Appendix \ref{bse_det}. Figure \ref{flow} presents the flowchart of the calculations which gives a general idea of how the calculations are organized. The diagrammatic (GW and the diagramms beyond GW) parts of the FlapwMBPT code take full advantage of the fact that certain diagrams can more efficiently be evaluated in reciprocal (and frequency) space whereas other diagrams are easier to evaluate in real (and time) space. As a result, GW part of the code scales as $N_{k}N_{\omega}N^{3}_{b}$ where $N_{k}$ is the number of \textbf{k}-points in the Brillouin zone, $N_{\omega}$ is the number of Matsubara frequencies, and $N_{b}$ stands for the size of the basis set. The vertex part of the code scales as $N^{2}_{k}N^{2}_{\omega}N^{4}_{b}$. For comparison, if one uses naive (all in reciprocal space and frequency) implementation then GW part scales as $N^{2}_{k}N^{2}_{\omega}N^{4}_{b}$ (i.e. exactly as the vertex part when the implementation is efficient), and the vertex part scales as $N^{3}_{k}N^{3}_{\omega}N^{5}_{b}$. Besides of efficiency of the implementation, we have to mention two more factors which make the use of the diagrams beyond GW feasible. First is the fact that the higher order diagrams converge much faster than the GW diagram with respect to the basis set size and to the number of \textbf{k}-points.\cite{prb_94_155101,prb_95_195120} Second is that the higher order diagrams are very well suited for massive parallelization.

We also use quasiparticle self consistent GW (QSGW) approach. Similar to scGW, sc(GW+G3W2), and sc(BSE:P$@$GW+G3W2) approaches, it is based on the finite temperature (Matsubara) formalism and in this respect it is different from the well known QSGW implementation by Kotani et al.\cite{prb_76_165106} Quasiparticle approximation includes linearization of self energy near the zero frequency (see for details Refs. [\onlinecite{prb_85_155129,cpc_219_407}]) and, therefore, the method is reliable only not very far from the Fermi level - usually within a few electron-volts. Effective self energy is static (frequency independent) and the method is not diagrammatic. However, as it was explained by Kotani et al.\cite{prb_76_165106}, QSGW satisfies the zero frequency and long wave limit of the Ward Identity (WI) because of the so called Z-factor cancellation. This fact makes it often quite accurate, especially in simple metals and semiconductors where the above mentioned limit is important. Considering the differences between QSGW and the above introduced approaches, together they represent a good set of methods to study new materials.

Principal difference between fully relativistic calculations (FR) ans scalar relativistic (SR) calculations consists in the fact that we use Dirac-Kohn-Sham equations to generate LAPW+LO basis set in the FR case (see Ref. [\onlinecite{prb_103_165101}] for the implementation in the FlapwMBPT code) instead of scalar-relativistic Kohn-Sham equations.\cite{zpb_32_43} Generalization of the evaluation of diagrams to the FR case is relatively straightforward: one just replaces the SR basis functions with FR basis functions in the evaluation of matrix elements (see for instance the generalization of scGW and QSGW to fully relativistic variant in Ref. [\onlinecite{prb_85_155129}]).

Let us now specify the setup parameters used in the calculations. In order to make presentation more compact, principal structural parameters for the studied solids have been collected in Table \ref{list_s} and the most important set up parameters have been collected in Table \ref{setup_s}. All calculations have been performed for the electronic temperature $600K$. In all calculations we assumed the ferromagnetic (FM) ordering. The calculations (excluding the vertex part) were performed with the $4\times 4\times 4$ mesh of \textbf{k}-points in the Brillouin zone. 500 band states (1000 in the FR case) were used to expand Green's function and self energy. Product basis (PB) consisted of approximately 3100 functions (depending on \textbf{k}-point). The diagrams beyond GW approximation were evaluated using $2\times 2\times 2$ mesh of \textbf{k}-points in the Brillouin zone and with about 40 (80 in the FR case) bands (closest to the Fermi level). With the above mentioned faster convergence of higher order diagrams with respect to these parameters, this choice represented a reasonable compromise between the accuracy and the computational cost. Most important convergence tests are presented in Tables \ref{gap_helo}, \ref{conv_k}, and \ref{conv_vrt}. As one can deduce from the convergence tests, the remaining uncertainty of the band gap obtained in fully relativistic sc(BSE:P$@$GW+G3W2) calculations could be at the level of 0.1--0.2 eV. Also, most likely effect of further refining of the computational setup would be a reduction of the calculated band gap.

\section*{Results}
\label{res}

\begin{table}[t]
\caption{Structural parameters of the solids studied in this work. Lattice parameters are in Angstroms, MT radii are in atomic units (1 Bohr radius), and atomic positions are given relative to the three primitive translation vectors. Experimental structural data from Ref. [\onlinecite{cm_27_612}] are used.} \label{list_s}
\small
\begin{center}
\begin{tabular}{@{}c c c c c c} &Space&&&Atomic&\\
Solid &group&a&c&positions&$R_{MT}$\\
\hline\hline
CrI$_{3}$&148 &6.867 &19.807&Cr: 1/3;2/3;0.33299  &2.471\\
& & & &I: 0.31677;0.33453;0.4123  &2.667\\
\end{tabular}
\end{center}
\end{table}

\begin{table}[t]
\caption{Principal setup parameters of the studied solids are given. The following abbreviations are introduced: $\Psi$ is for wave functions, $\rho$ is for the electronic density, $V$ is for Kohn-Sham potential, and PB is for the product basis.} \label{setup_s}
\small
\begin{center}
\begin{tabular}{@{}c c c c c c} &Core&&$L_{max}$&$L_{max}$&\\
Solid &states&Semicore&$\Psi/\rho,V$&PB & $RK_{max}$ \\
\hline\hline
CrI$_{3}$&Cr: [Ne]& 3s,3p&6/6&6&6.0  \\
& I: [Kr]& 5s,4d&6/6&6&  \\
\end{tabular}
\end{center}
\end{table}

\begin{table}[t]
\begin{center}
\caption{Convergence of the band gaps obtained in scalar relativistic G0W0 calculations with respect to the number of high energy local orbitals (HELO) included in the LAPW+LO basis set. Local orbitals associated with semicore states are not included. Numbers after orbital character indicate how many LO's are included with a given orbital character. The results presented in the main text correspond to the second row (i.e. s2p1d2/s1p2d1).} \label{gap_helo}
\begin{tabular}{@{}c c c}  \multicolumn{2}{c}{High energy LO} & \\
Cr  &I & Band gap (eV)\\
\hline\hline
s1d1 & p1 & 2.09 \\
s2p1d2 & s1p2d1 & 2.07 \\
s2p1d2f1 & s1p2d1f1 & 2.07  \\
s3p2d3f2 & s2p3d2f2 & 2.09  \\
s3p3d4f3 & s3p4d3f3 & 2.10  \\
\end{tabular}
\end{center}
\end{table}

\begin{table}[t]
\caption{Dependence of the calculated band gap of CrI$_{3}$ on the \textbf{k}-grid $N_{\mathbf{k}}$ in G0W0 calculations. Scalar relativistic approach has been used.} \label{conv_k}
\begin{center}
\begin{tabular}{@{}c c c}  $N_{\mathbf{k}}$ & Band gap\\
\hline\hline
$2^{3}$ &2.31 \\
$3^{3}$ &2.16 \\
$4^{3}$ & 2.07\\
$5^{3}$ &2.09 \\
$6^{3}$ & 2.06\\
\end{tabular}
\end{center}
\end{table}

\begin{table}[t]
\caption{Dependence of the calculated band gap of CrI$_{3}$ on the calculation setup for the diagrams beyond $GW$. Scalar relativistic sc(GW+G3W2) approach has been used. $N^{vrt}_{bnd}$ means the number of band states included in the evaluation of the beyond-GW diagrams. $N^{vrt}_{\mathbf{k}}$ means the \textbf{k}-grid used for the evaluation of the beyond-GW diagrams. Dependence on the $N^{vrt}_{bnd}$ was studied with fixed grid of \textbf{k}-points: $4\times 4\times 4$ for GW part and $2\times 2\times 2$ for vertex part. Dependence on the $N^{vrt}_{\mathbf{k}}$ was studied with fixed grid of \textbf{k}-points $6\times 6\times 6$ for GW part and with $N^{vrt}_{bnd}=40$. "Saturation" of the band gap when $N^{vrt}_{bnd}$ reaches 40 is related to the fact that all important band states, i.e. Cr 3d and I 5p bands, are included.} \label{conv_vrt}
\begin{center}
\begin{tabular}{@{}c c c}  Parameter & Setup & Band gap\\
\hline\hline
$N^{vrt}_{bnd}$ & 20 &2.91 \\
 & 30 &2.72\\
 & 40 &2.25 \\
 & 50 &2.19 \\
 & 60 &2.16 \\
\hline
$N^{vrt}_{\mathbf{k}}$ & $1^{3}$ &2.49 \\
 & $2^{3}$ &2.25 \\
 & $3^{3}$ & 2.27\\
\end{tabular}
\end{center}
\end{table}

\begin{table}[t]
\begin{center}
\caption{Band gaps (eV) and magnetic moments ($\mu_{B}$, Chromium site) of CrI$_{3}$ obtained at different levels of theory. SR stands for scalar-relativistic approximation, and FR is for fully relativistic approach. The positions of the peaks in \textbf{k}-resolved spectral functions have been used to measure the band gaps. This is demonstarted in Fig. \ref{dosk}. Two variants of G0W0 differ by starting point: Perdew–Burke–Ernzerhof (PBE) functional\cite{prl_77_3865} and Hartree-Fock (HF) approximation.} \label{gap_cri3}
\small
\begin{tabular}{@{}c c c c c}  &\multicolumn{2}{c}{Band gap} & \multicolumn{2}{c}{Moment}\\
Approximation  &SR & FR  &SR & FR\\
\hline\hline
LDA &0.85 &0.66  &2.95 & 3.06  \\
G0W0(PBE) &2.07 &1.99  &NA & NA  \\
G0W0(HF) &4.22 &3.74  &NA &NA \\
QSGW &3.11 &2.64  &3.08 &3.11 \\
scGW &3.03 &2.51  &3.23 &3.35   \\
sc(GW+G3W2) &2.25 &1.97  &3.21 &3.32 \\
sc(BSE:P@GW+G3W2) &1.86 &1.57  &3.20 &3.31   \\
\hline
Experiment:   && & & \\
Optical gap [\onlinecite{jap_36_1259}]   &\multicolumn{2}{c}{1.24} & \multicolumn{2}{c}{}\\
ARPES [\onlinecite{scirep_10_15602}]   &\multicolumn{2}{c}{1.3} & \multicolumn{2}{c}{}\\
\end{tabular}
\end{center}
\end{table}

\begin{figure}[t] 
\begin{center}       
    \includegraphics[width=6.5 cm]{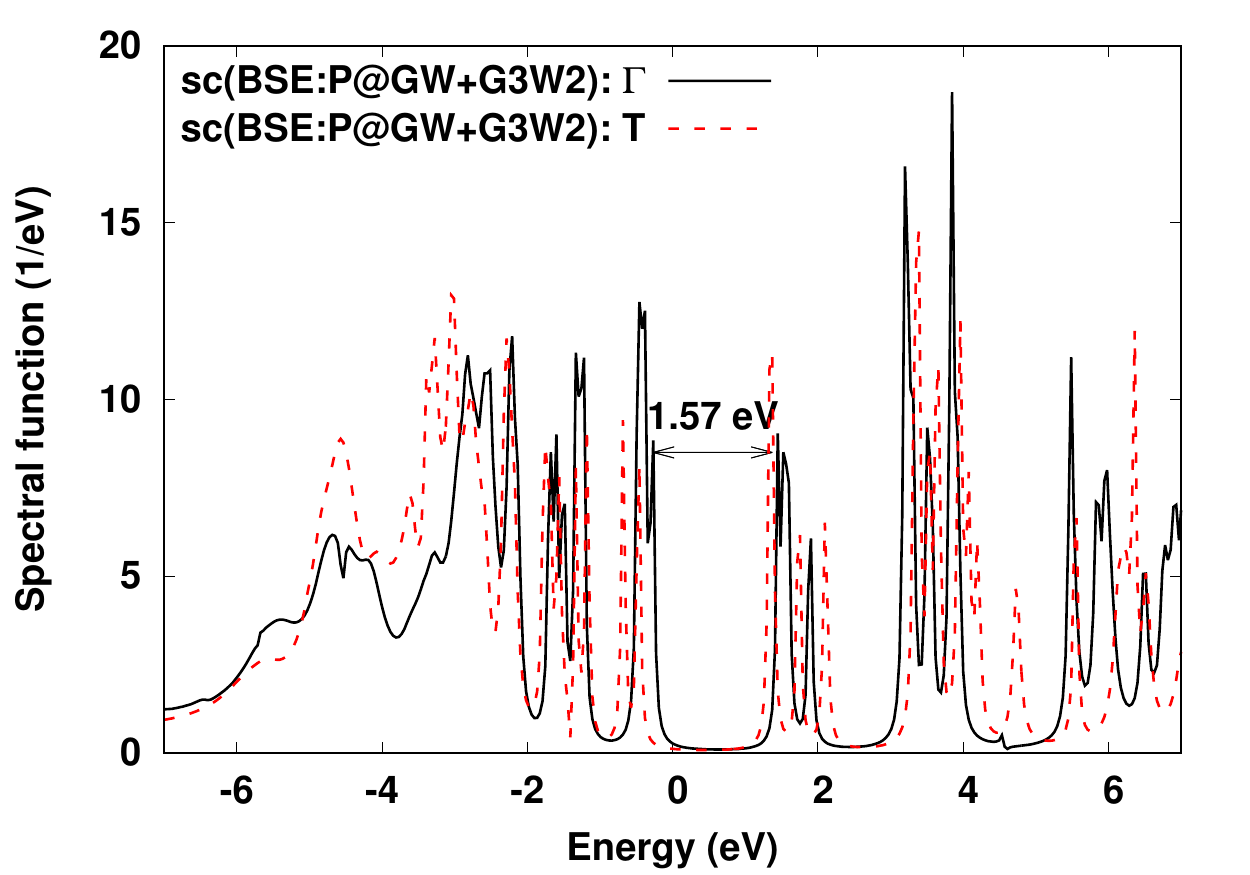}
    \caption{Spectral function of CrI$_{3}$ at $\Gamma$ and $T$ points in the Brillouin zone as obtained in fully relativistic sc(BSE:P$@$GW+G3W2) approach. The value of the band gap defined as the difference in the positions of peaks is shown.}
    \label{dosk}
\end{center}
\end{figure}

\begin{figure*}[t]       
    \fbox{\includegraphics[width=6.5 cm]{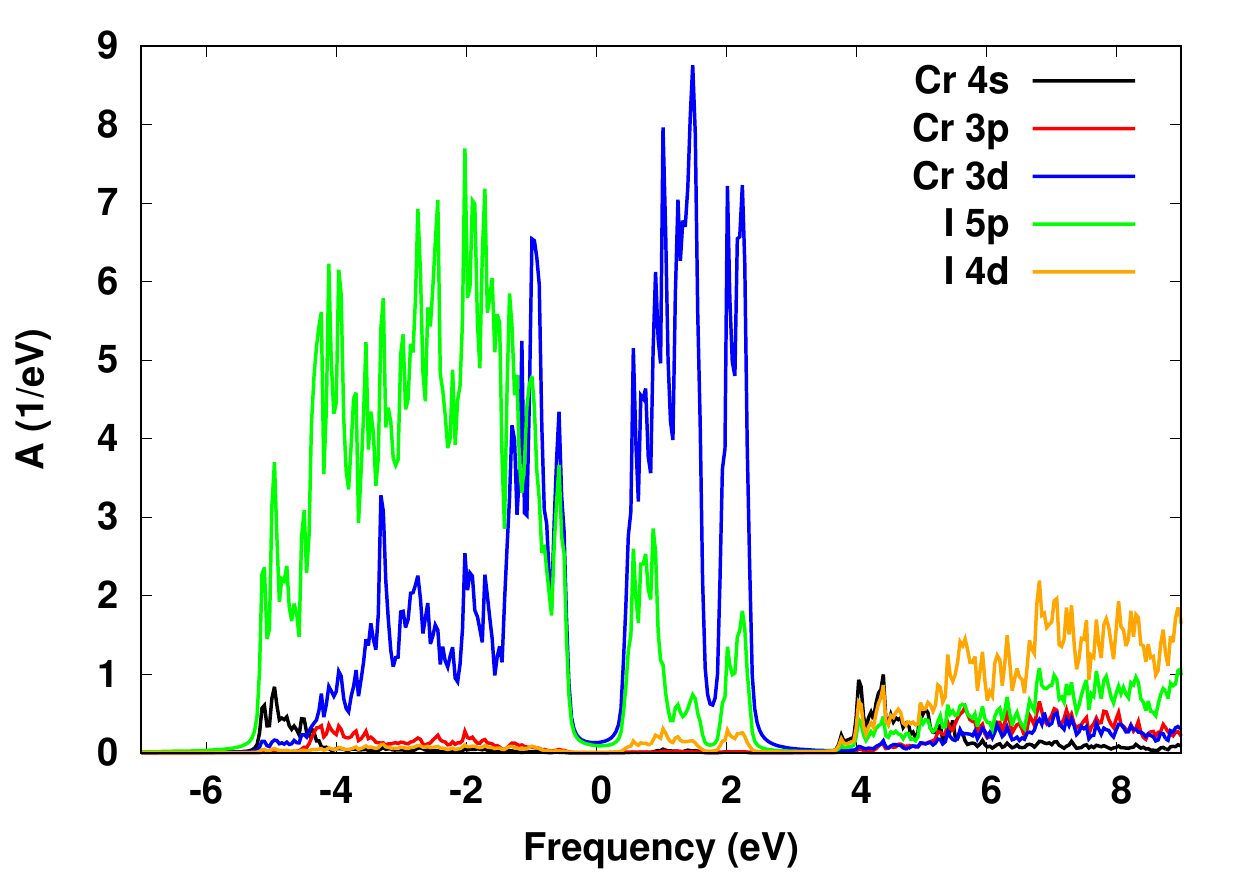}}   
    \hspace{0.02 cm}
    \fbox{\includegraphics[width=6.5 cm]{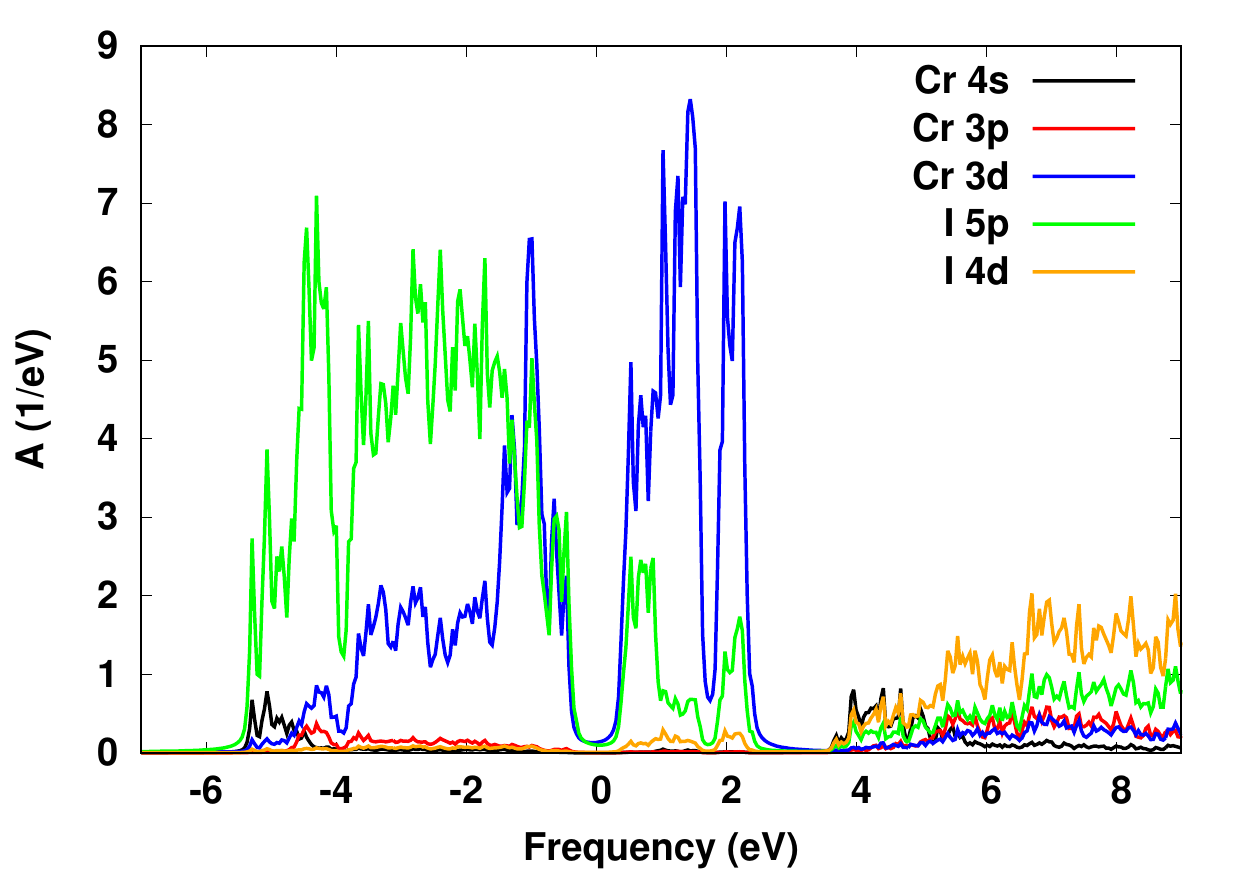}}  
    \hspace{0.02 cm}
    \caption{Total and partial (atom and orbital resolved) spectral functions of CrI$_{3}$ obtained in LDA calculations. Scalar relativistic results are in the left window. Fully relativistic results are in the right window. Sums of spin-up and spin-down quantities in the SR case, and sums of spin-orbit components (i.e. $p_{1/2}+p_{3/2}$ and $d_{3/2}+d_{5/2}$) in the FR case are given.}
    \label{pdos_dft}
\end{figure*}

We begin the presentation of results by showing in Table \ref{gap_cri3} the band gaps and magnetic moments (on chromium sites) obtained using different approximations. Magnetic moments do not show any noticeable dependence on the method and are in accordance with other calculations.\cite{njp_21_053012} They also depend slightly on the choice of the muffin-tin radii and, correspondingly, are given here just for the reference. Calculated band gaps, however, show remarkable dependence on the approximation used. As usual, LDA underestimates the band gap by about 30-50\% depending on how one approximates the relativistic effects. Both QSGW and scGW seriously overestimate the experimental band gap (by about factor of two). QSGW does not show improvement in the calculated band gap of CrI$_{3}$ as compared to scGW, which one would expect in small gap sp semiconductors.\cite{prb_98_155143} From this fact, one can conclude that the presence of Cr 3d electrons makes this material somewhat different from the simple semiconductors. Noticeable improvement in the evaluated band gap happens when we include first order vertex correction, i.e. when we switch from scGW to sc(GW+G3W2). Further improvement, i.e. when we switch from sc(GW+G3W2) to sc(BSE:P$@$GW+G3W2), is a bit smaller. The effect of inclusion/neglecting the SOC is approximately of the same amplitude as the effect of using BSE when we consider the SOC effect at sc(BSE:P$@$GW+G3W2) level. At this level it is about twice smaller than in Ref. [\onlinecite{prb_101_241409}] which means that the self-consistent inclusion of the SOC makes some difference. At the level of scGW/QSGW, however, the effect of SOC is somewhat larger. It is interesting that the best (and the most sophisticated) result for the band gap in our study (1.57 eV) is quite close to the result 1.68 eV obtained in [\onlinecite{prb_101_241409}] using empirical enhancement of the screening. Thus, if we assume that there is no big differences in QSGW between this study and Ref. [\onlinecite{prb_101_241409}], we can state that QSGW80 works rather well for this material.

\begin{figure*}[t]       
    \fbox{\includegraphics[width=6.5 cm]{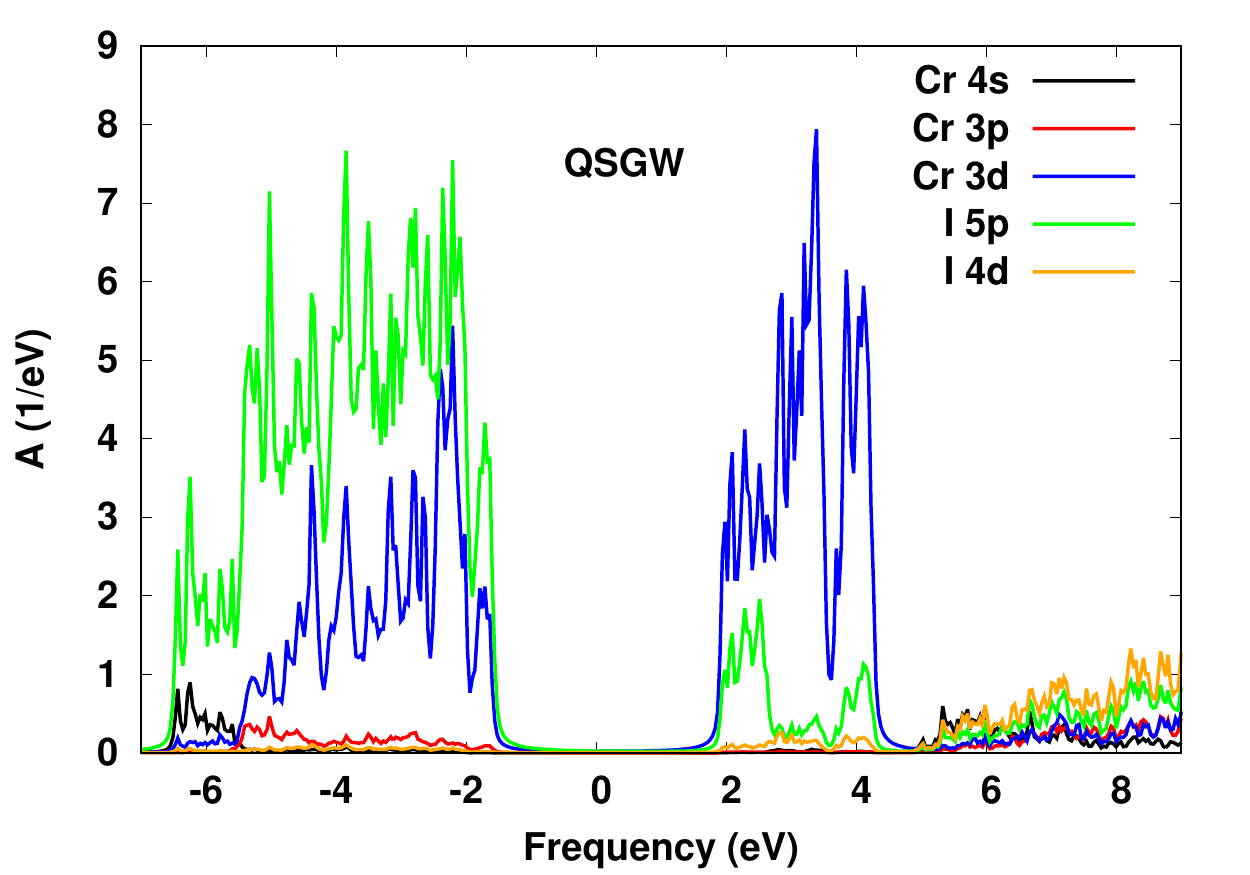}}   
    \hspace{0.02 cm}
    \fbox{\includegraphics[width=6.5 cm]{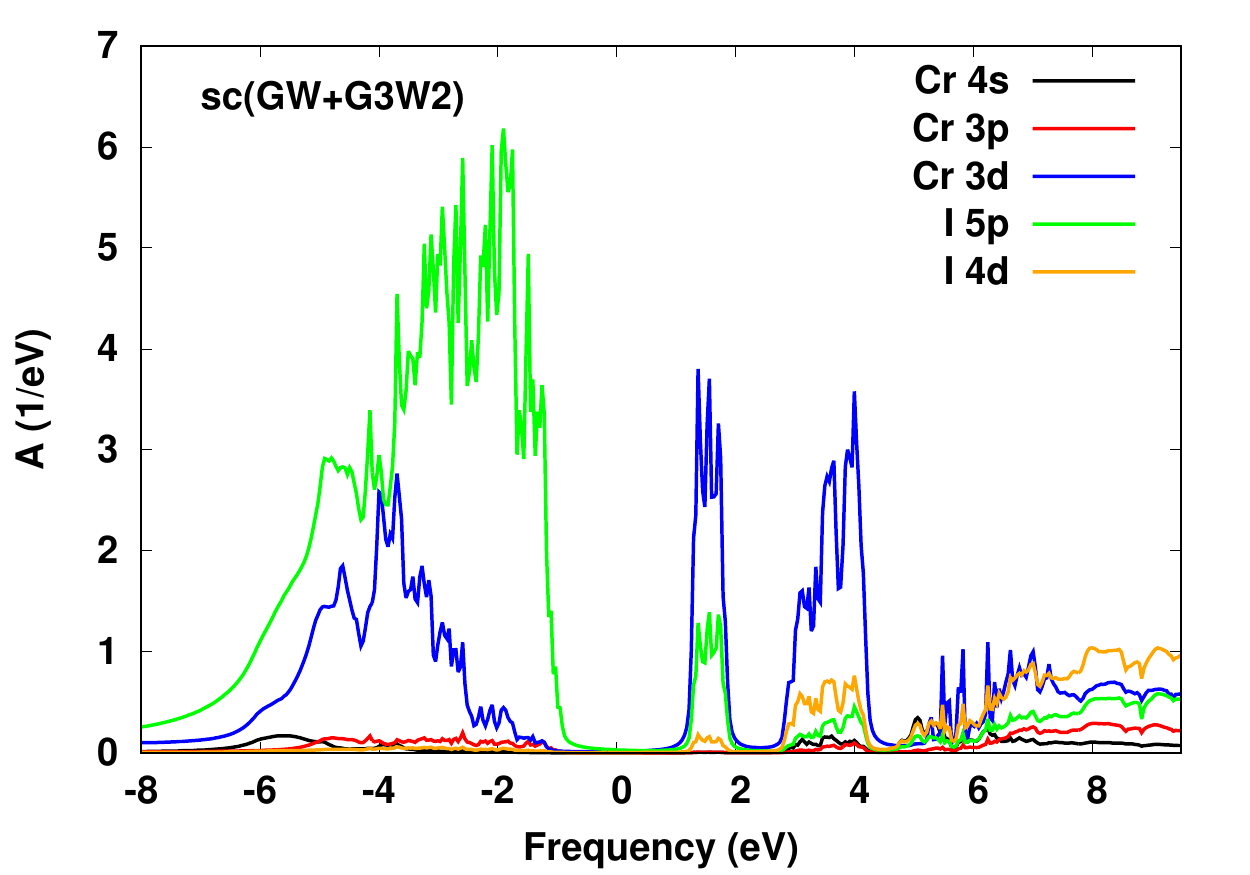}}  
    \hspace{0.02 cm}
    \fbox{\includegraphics[width=6.5 cm]{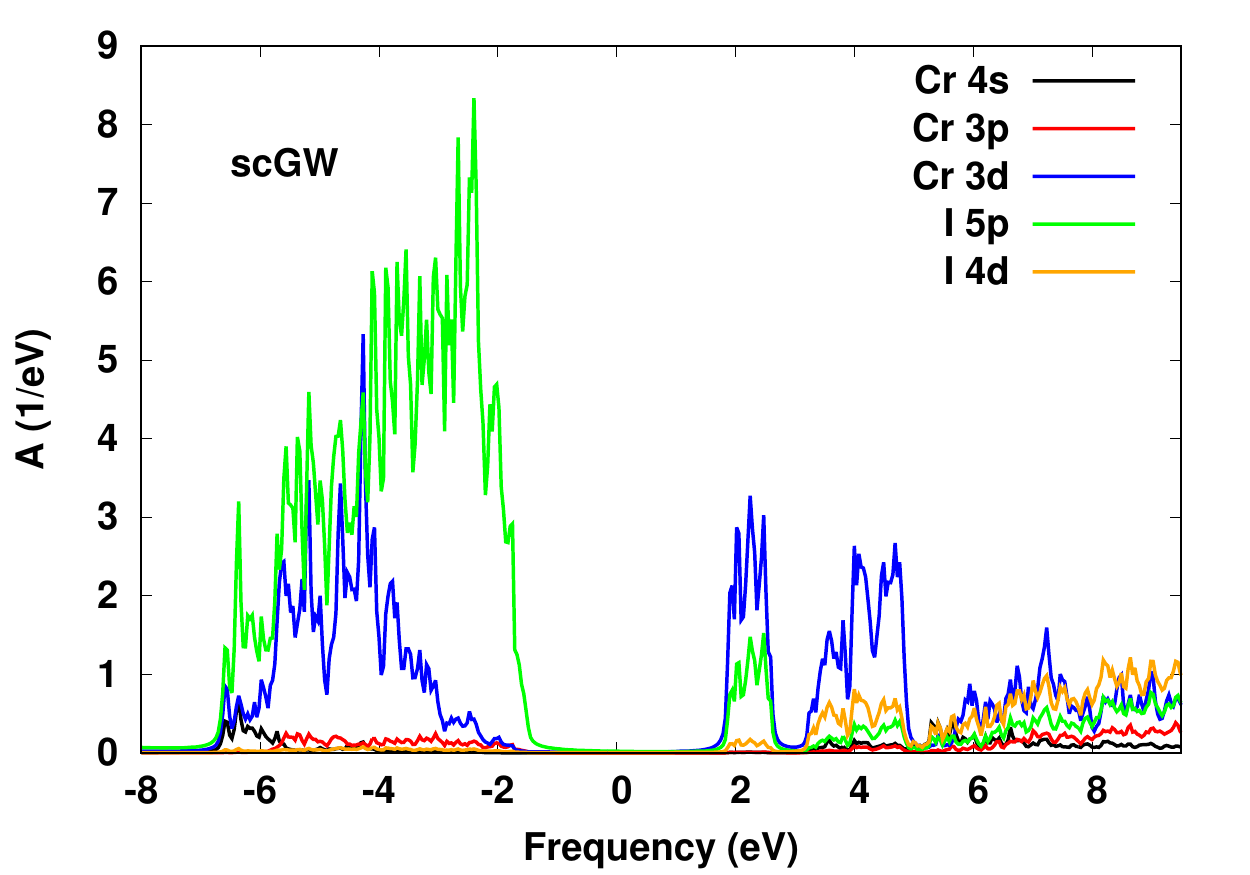}}  
    \hspace{0.02 cm}
    \fbox{\includegraphics[width=6.5 cm]{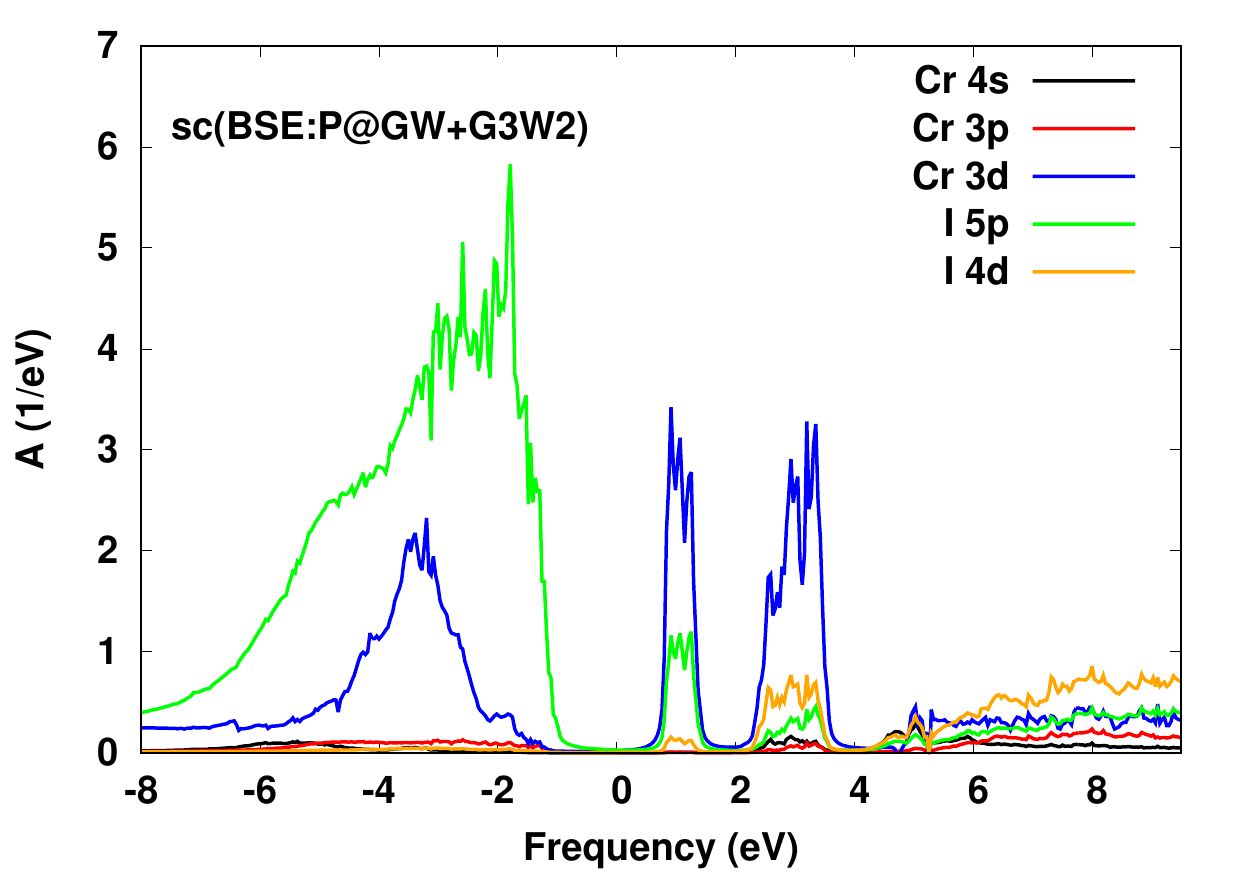}}
    \caption{Total and partial (atom and orbital resolved) spectral functions of CrI$_{3}$ obtained in Green's function based methods. Scalar relativistic results. Sums of spin-up and spin-down quantities are given. Analytical continuation of self energy\cite{jltp_29_179,cpc_257_107502} was used to get Green's function on the real frequency axis. The curves become smoother in the sequence QSGW-scGW-sc(GW+G3W2)-sc(BSE:P$@$GW+G3W2) primarily because of increase in the many-body effects (incoherence).}
    \label{pdos_sr}
\end{figure*}

\begin{figure*}[t]       
    \fbox{\includegraphics[width=6.5 cm]{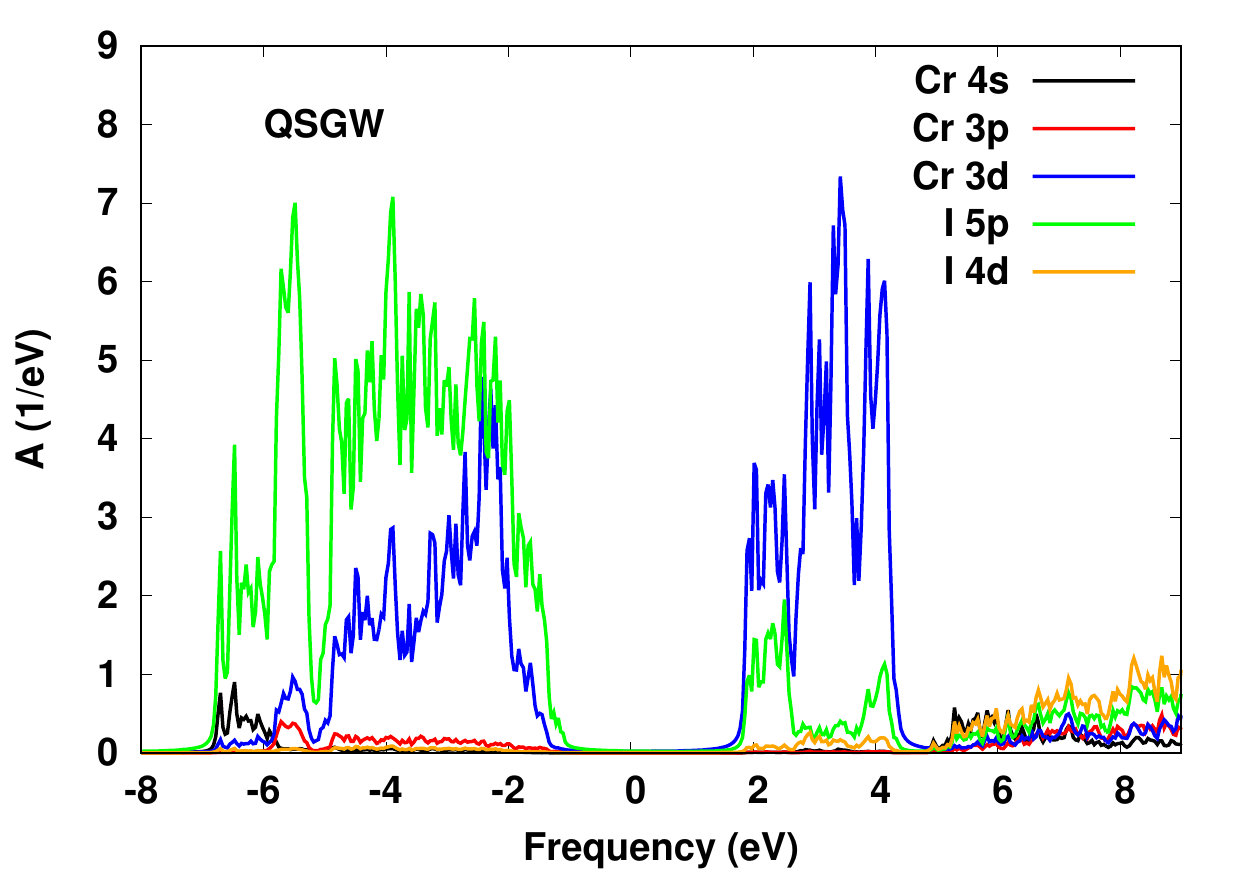}}   
    \hspace{0.02 cm}
    \fbox{\includegraphics[width=6.5 cm]{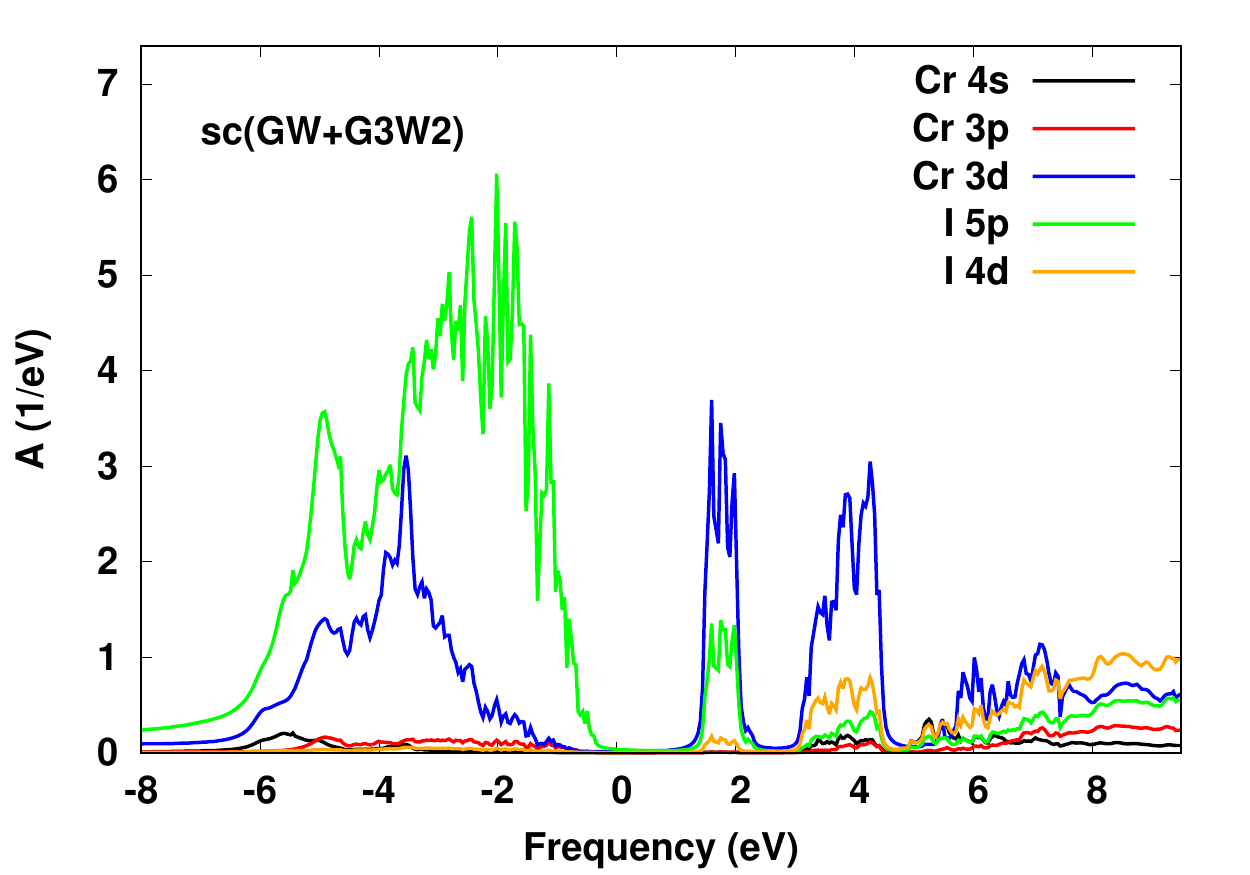}}  
    \hspace{0.02 cm}
    \fbox{\includegraphics[width=6.5 cm]{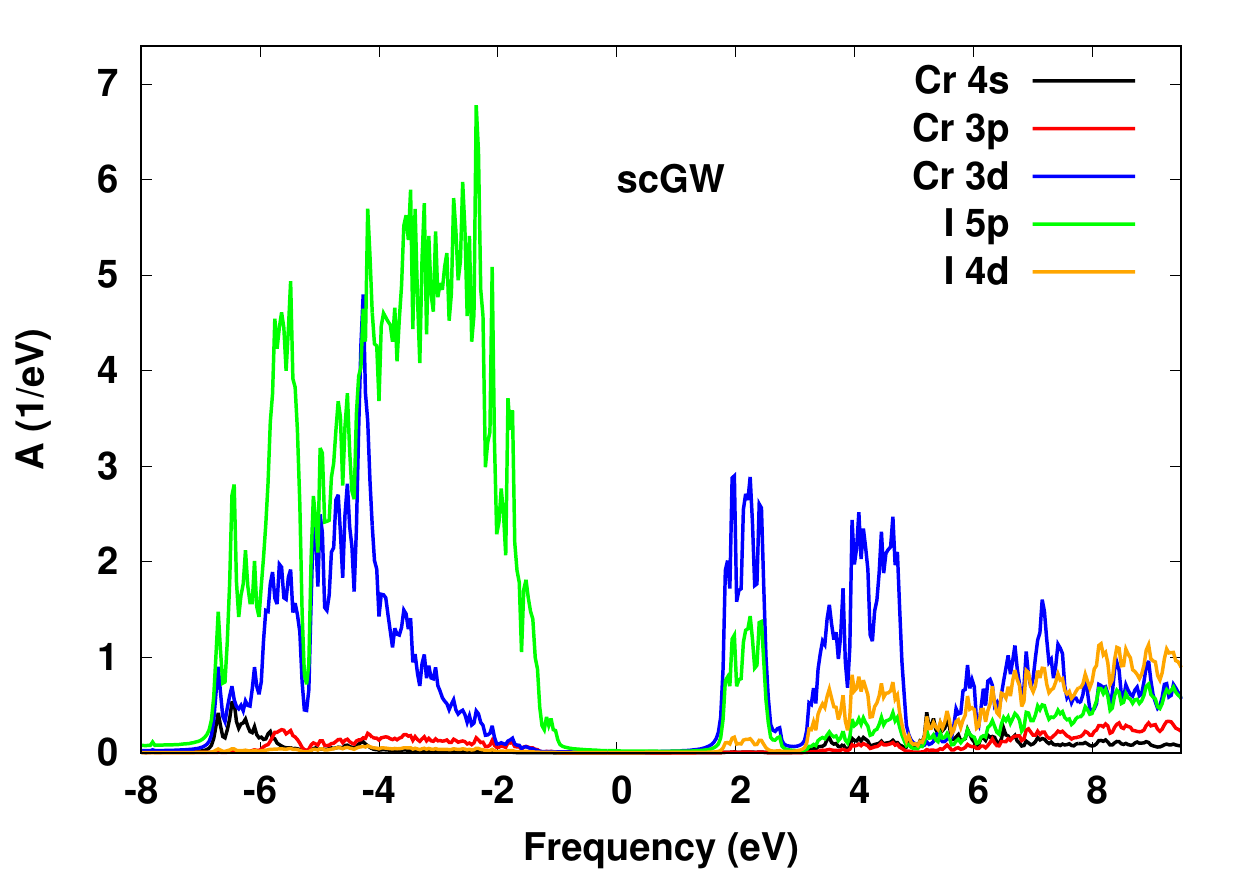}}  
    \hspace{0.02 cm}
    \fbox{\includegraphics[width=6.5 cm]{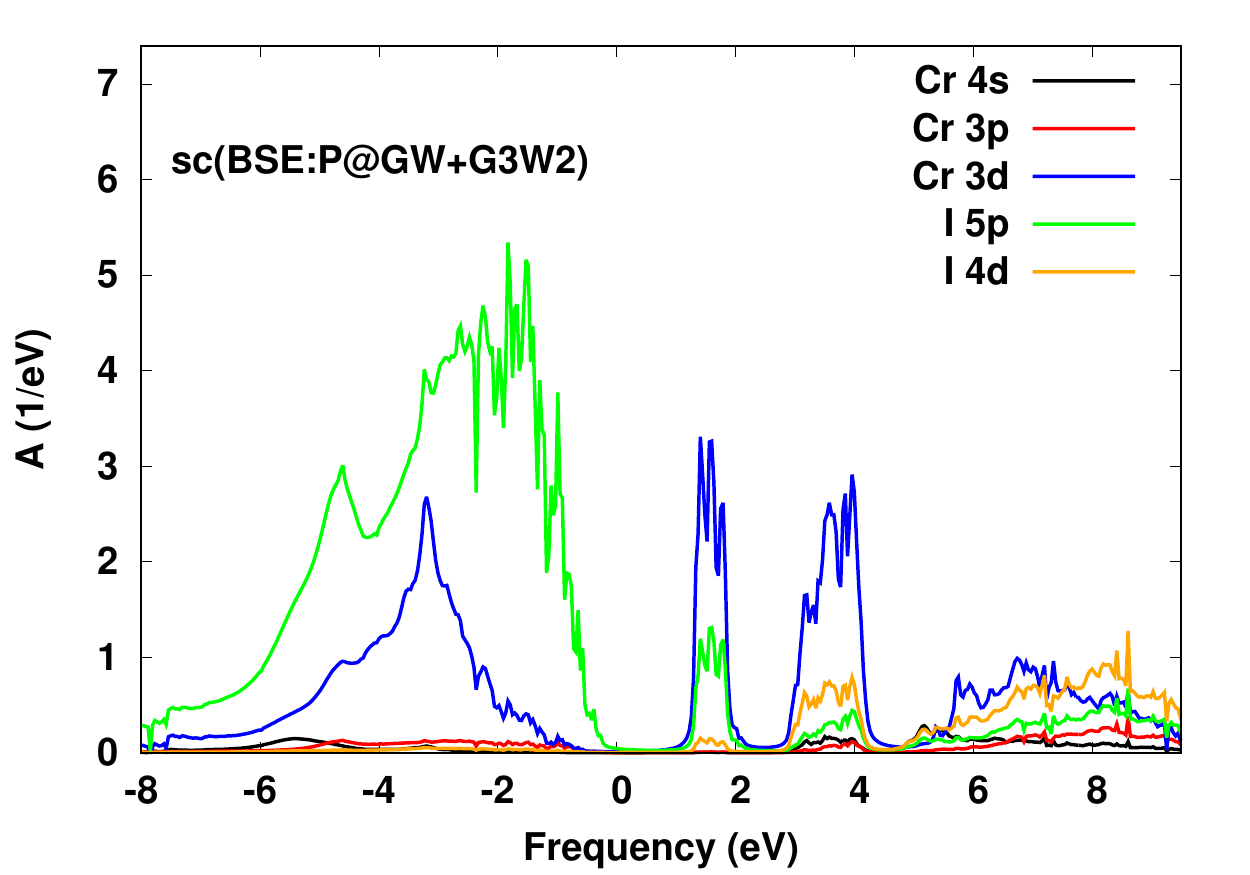}}
    \caption{Total and partial (atom and orbital resolved) spectral functions of CrI$_{3}$ obtained in Green's function based methods. Fully relativistic results.  Sums of spin-orbit components (i.e. $p_{1/2}+p_{3/2}$ and $d_{3/2}+d_{5/2}$) are given. Analytical continuation of self energy\cite{jltp_29_179,cpc_257_107502} was used to get Green's function on the real frequency axis.}
    \label{pdos_fr}
\end{figure*}

Our final result for the band gap (1.57 eV) still is a bit larger as compared to the experimental 1.3 eV obtained in ARPES studies.\cite{scirep_10_15602} One can name a few possible reasons for this remaining disagreement: i)  numerical cutoffs (especially in the vertex part); ii)higher order diagrams not included in this study; iii) electroh-phonon interaction. All three reasons, normally, should result in some reduction of the calculated band gap bringing it in even better agreement with the experiment. But even at the present level, the error is already small enough and allows us to state that this material is a weakly correlated one and can be described using ab-initio diagrammatic methods. 

In Fig. \ref{pdos_dft} we show partial density of states (atom and orbital resolved) of CrI$_{3}$ obtained in LDA calculations. Besides a little shrinkage of the band gap in fully relativistic case, there is very little difference between scalar relativistic and fully relativistic results. As one can see, principal spectral features around the Fermi level are almost completely defined by Cr 3d and I 5p states. In this respect, one can point out to a certain disagreement with the experimental ARPES data obtained by Kundu et al.\cite{scirep_10_15602} Namely, in experiments, valence band maximum (VBM) is formed by I 5p states only and Cr 3d states are shifted downward by about 0.6 eV. However, there is no such separation between I 5p and Cr 3d states in LDA calculations. Thus, we can conclude that LDA not only underestimates the band gap by almost 50\% but also predicts incorrect distribution of the orbital character among the valence bands.

In Fig. \ref{pdos_sr} we present partial spectral functions for Green's function based methods as obtained in scalar relativistic approximation. Similar results obtained in fully relativistic approach are shown in Fig. \ref{pdos_fr}. Similar to the DFT case, there is no considerable difference between SR and FR results. So, our discussion is relevant to both figures equally. Firstly, we point out that QSGW approximation does not show a shift between Cr 3d and I 5p states. In this respect it is in a disagreement with the ARPES (as LDA is). Its difference with LDA is only in the considerable overestimation of the band gap. The rest of methods (scGW, sc(GW+G3W2), and sc(BSE:P$@$GW+G3W2)) clearly show the separation between Cr 3d and I 5p states. In these three methods VBM is formed solely by I 5p orbitals (as in experiments) and the onset of Cr 3d states is shifted downward from the VBM by 0.5--1.0 eV in agreement with the separation 0.6 eV found in the ARPES measurements.\cite{scirep_10_15602} The difference between QSGW and other three methods is that self energy is static (frequency independent) in QSGW whereas three other methods take full frequency dependence of self energy into account. Obviously, this frequency dependence is crucial for CrI$_{3}$. Another qualitative feature missing in QSGW consists in breaking the Cr 3d states in the conduction bands into two groups. Figures \ref{pdos_sr} and \ref{pdos_fr} also show gradual reduction of the band gap, but this was already discussed above.

Important comment about second order (in W) vertex correction to self energy has to be given. The problem of negative spectral weight appearance (when one uses this correction) was discussed and certain measures were taken to circumvent the issue.\cite{prb_90_115134,prb_91_115104,prb_102_045121} Particularly, it was stated that it is impossible to perform self-consistent calculations which include G3W2 correction.\cite{prb_90_115134} However, as it appears, sc(GW+G3W2) calculations can definitely be performed for CrI$_{3}$. They were also performed for a number of other systems\cite{prb_95_195120,arx_2105_03770,arx_2106_03800} and also for electron gas\cite{prb_96_035108} where sufficiently high convergence can be achieved. Besides of considerable increase in computer time needed, sc(GW+G3W2) calculations did not show any additional problems as compared to scGW calculations. Author of this work does not know the explanation of why the issue does not reveal itself. May be the reason is that all sc(GW+G3W2) (as well as scGW) calculations are performed using Matsubara's frequency axis and this fact somehow conceals the problem. Or, may be the self consistence itself, in fact, cures the problem because the sc(GW+G3W2) approach is $\Psi$-derivable and therefore is conserving.

As it follows from the above discussion, basic features of the electronic structure known from experiments (the band gap and Cr 3d/I 5p separation) can quite accurately be described using ab-initio diagrammatic methods. Thus, there is no need to apply the methods with adjustable parameters (DFT+U or DFT+DMFT) to study CrI$_{3}$ and, most likely, other materials from this class.

\section*{Nonlocal effects}
\label{nonloc}

\begin{figure*}[t]       
    \fbox{\includegraphics[width=6.5 cm]{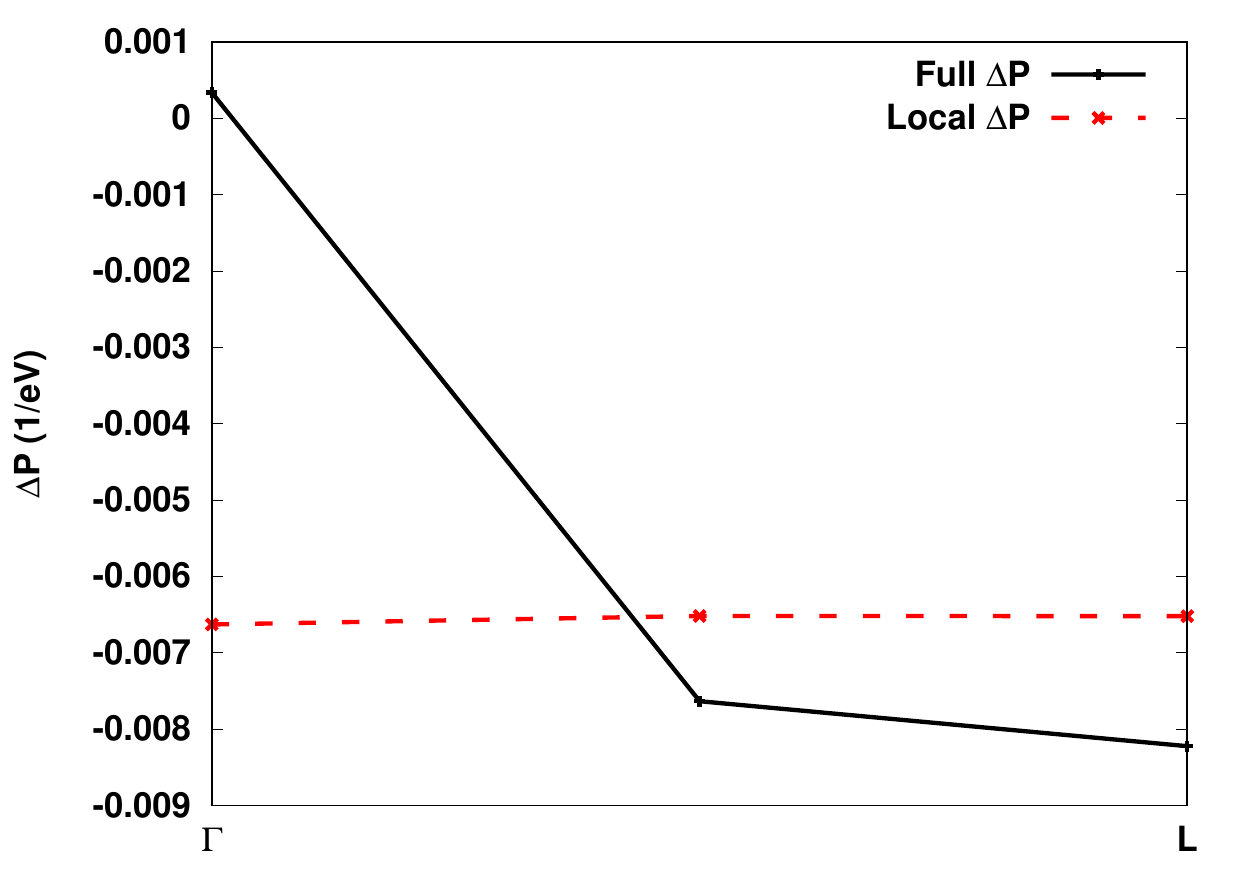}}   
    \hspace{0.02 cm}
    \fbox{\includegraphics[width=6.5 cm]{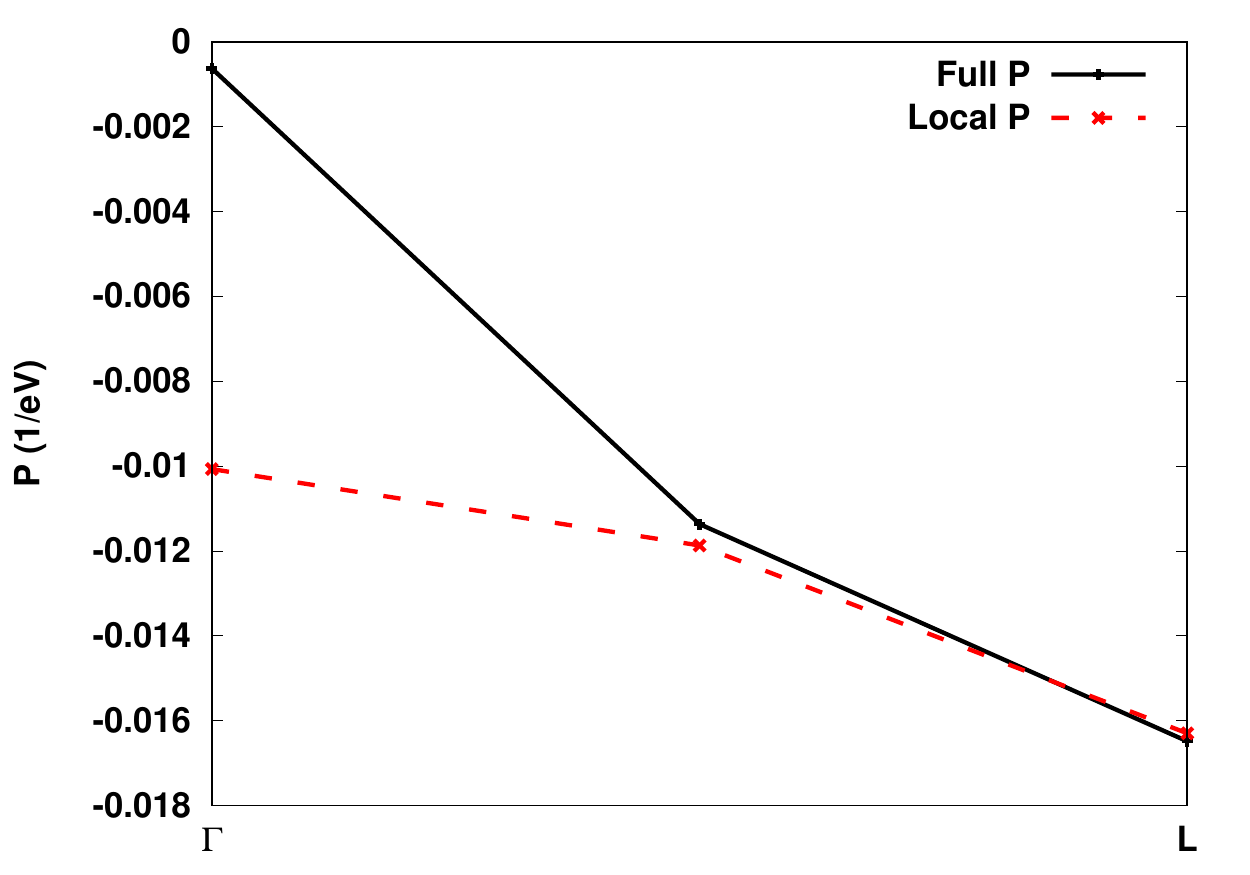}}  
    \hspace{0.02 cm}
    \caption{The components $P^{\mathbf{q}}_{\mathbf{G}=\mathbf{G}'=0}(\nu=0)$ of the calculated irreducible polarizability as functions of the momentum $\mathbf{q}$ along the direction $\Gamma$-L in the Brillouin zone. Vectors \textbf{G} and \textbf{G}' represent reciprocal lattice translations. Left window shows the vertex correction, and in the right window one can see the full polarizability.}
    \label{p_q}
\end{figure*}

\begin{figure*}[t]       
    \fbox{\includegraphics[width=6.5 cm]{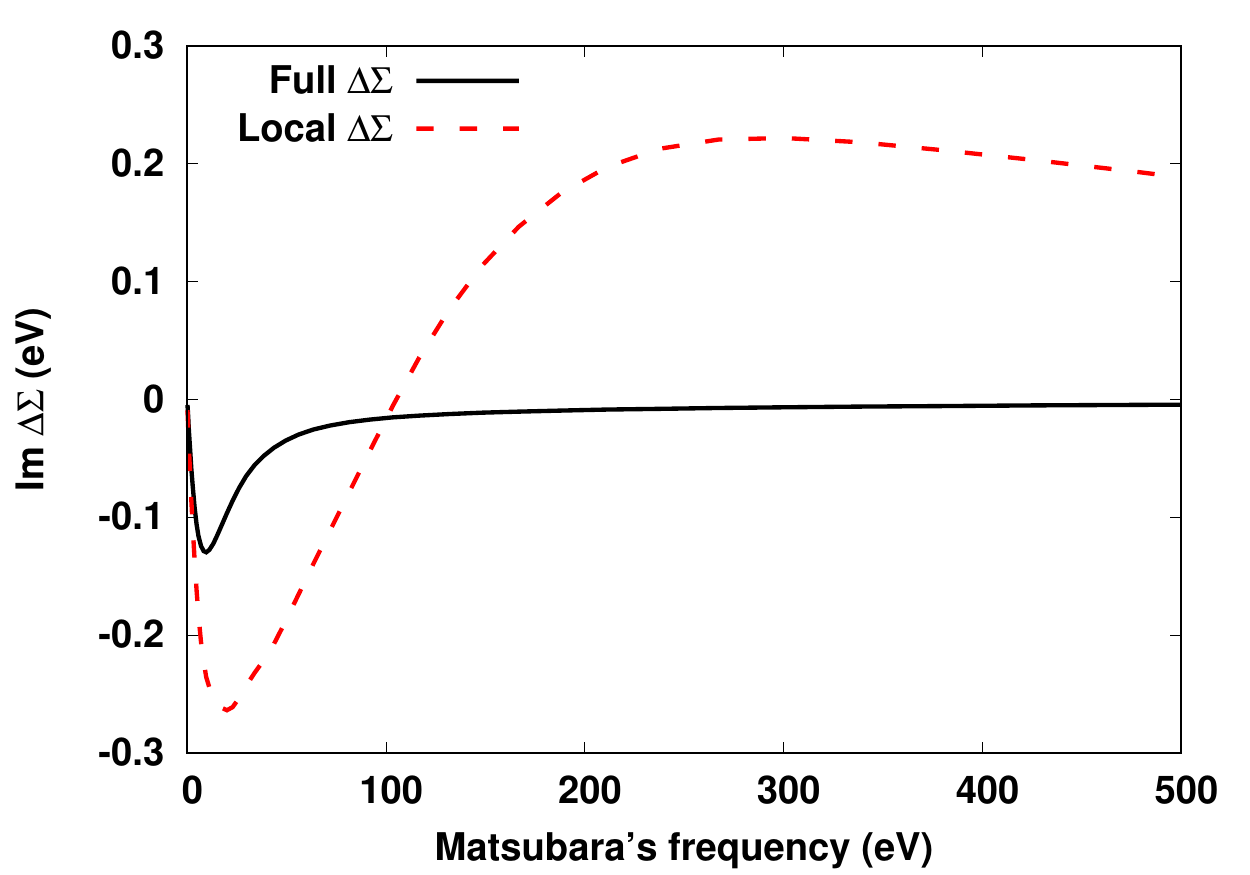}}   
    \hspace{0.02 cm}
    \fbox{\includegraphics[width=6.5 cm]{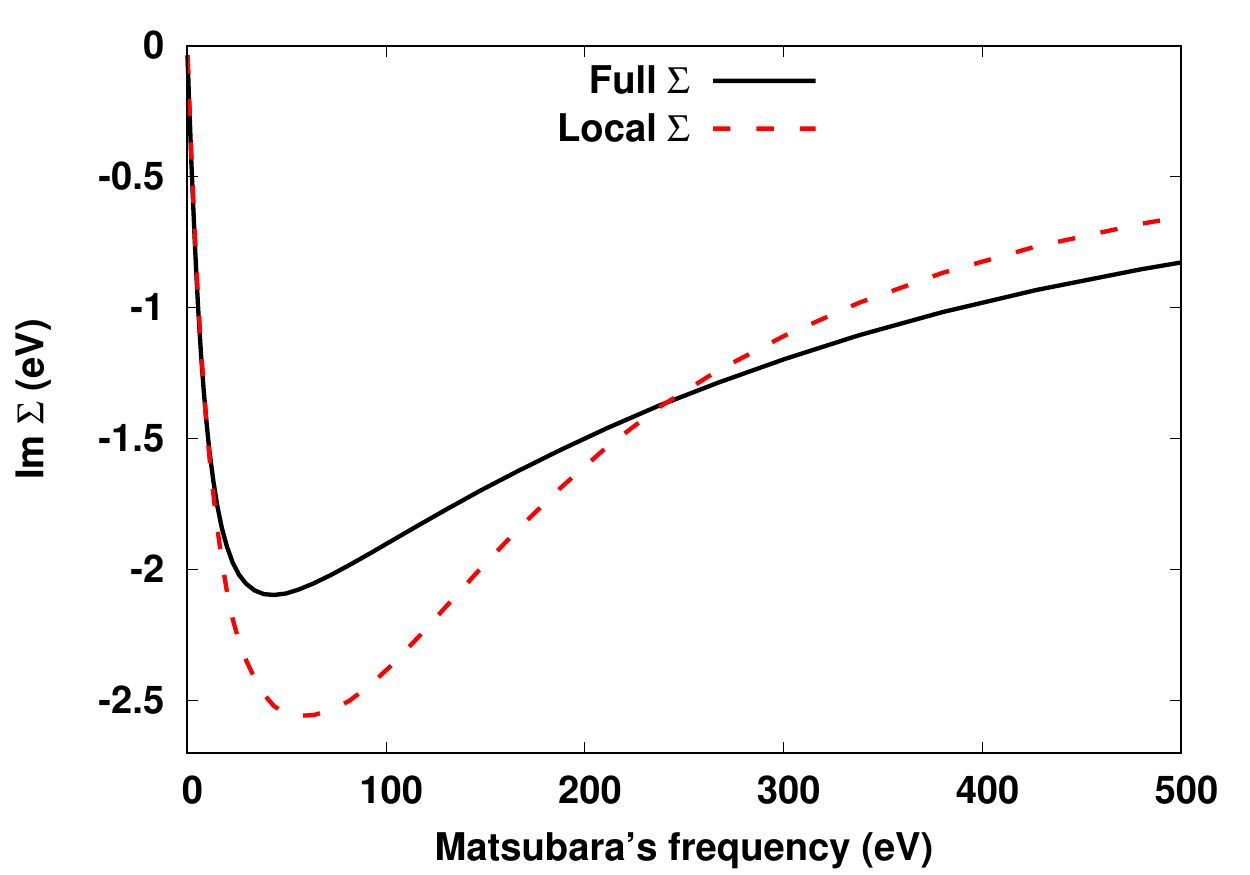}}  
    \hspace{0.02 cm}
    \caption{Imaginary part of self energy at $k=(0;0;0)$ as a function of Matsubara's frequency. Diagonal matrix element for the VBM band is used for plotting. Left window shows the vertex correction, and in the right window one can see full self energy.}
    \label{sig}
\end{figure*}

In order to check the quality of the local (single site) approximation we also performed simplified calculations at sc(GW+G3W2) level (scalar relativistic) and compared the results with the corresponding calculations which, however, take full non-locality into account. Instead of the \textbf{k}-dependent band states as a basis set in full calculations, we used a set of orbitals confined inside their muffin tin spheres as a basis in our simplified calculations. We have to point out that our simplified (single site) basis set was still slightly extended as compared to what normally would be used in, for instance, GW+DMFT study. Namely, for Cr sites, we included in the basis set not only 3d orbitals but also their energy derivatives as they naturally appear in the linearized augmented plane wave (LAPW) method. We also included 5p and their energy derivatives in the basis set on I sites. Single site approximation makes drastic effect on the performance: vertex corrections in this case take practically zero time to be evaluated. However, as we discuss below, the calculations performed with the single site approximation are not free from some issues.

Quite predictably, the most problematic for the local approximation quantity is the "head" of polarizability $P^{\mathbf{q}}_{\mathbf{G}=\mathbf{G}'=0}$, where vectors \textbf{G} and \textbf{G}' represent reciprocal lattice translations. Polarizability is an intrinsically non-local function in real space. In reciprocal space, the momentum dependence of its "head" at small momenta is $P^{\mathbf{q}}_{\mathbf{G}=\mathbf{G}'=0}=Bq^{2}$ in exact theory. This behavior cancels the $1/q^{2}$ divergence of the bare Coulomb potential at small momenta. In self consistent diagrammatic approaches we normally have $P^{\mathbf{q}}_{\mathbf{G}=\mathbf{G}'=0}=A+Bq^{2}$ with A being small and negative. Its absolute value is normally much smaller than the absolute value of the "head" at all \textbf{q}-points on our \textbf{q}-mesh with non-zero momenta. In practice, we evaluate (by fitting) the coefficients A and B and use only the $Bq^{2}$ part to proceed. The A coefficient becomes smaller when the number of the diagrams is increased (order by order or by using the BSE). To a certain degree its value also depends on the numerical approximations (cutoffs) within the same diagrammatic approach. In this respect, it is important to use \textbf{q}-dependent functions in the evaluation of polarizability. If, however, we accept the local approximation for the vertex part, the "head" of the correction to polarizability becomes momentum independent with very large A coefficient for total polarizability.

Figure \ref{p_q} illustrates the above discussion. In the full calculation, the "head" is slightly positive at $q=0$ which is to compensate the negative value obtained from the first diagram in Fig. \ref{diag_P} (GG part). As one can see from the right window of Fig. \ref{p_q} where the "head" of total polarizability is shown, the compensation is not complete because of the numerical approximations and the limited number of diagrams. The correction to the "head" of polarizability obtained in local approximation is essentially a constant (momentum independent) and it looks as if it approximates the average over the Brillouin zone value. It is large compared to the GG part which makes total polarizability a poor approximation to the correct function.

Another important function for comparison is self energy. An example of it for the VBM is shown in Fig. \ref{sig}. In the full calculation, the effects of interference make the vertex correction to self energy relatively small and very well localized in frequency space. It approximates zero when frequency is about 100 eV. The vertex correction to self energy obtained in local approximation looks quite differently. It is larger in absolute value and it is very slowly decaying function in frequency space. One can speculate that slow diminishing of the amplitude of self energy (local approximation) at high frequencies is somehow related to the truncation of screened interaction W. Truncation of W is most dangerous at high frequencies when it approaches bare Coulomb interaction and, therefore, is of long ranged nature. Thus, at least for CrI$_{3}$, the interference effects which are neglected in local approximation are quite important. Total self energy (right window in Fig. \ref{sig}) shows that differences in the vertex correction part make the total functions also quite different. It is important to point out that the difference in total self energy is a combined effect of the difference in vertex correction to self energy and the self-consistency effect which affects also the GW part of it.

\begin{figure}[t]       
    \includegraphics[width=6.5 cm]{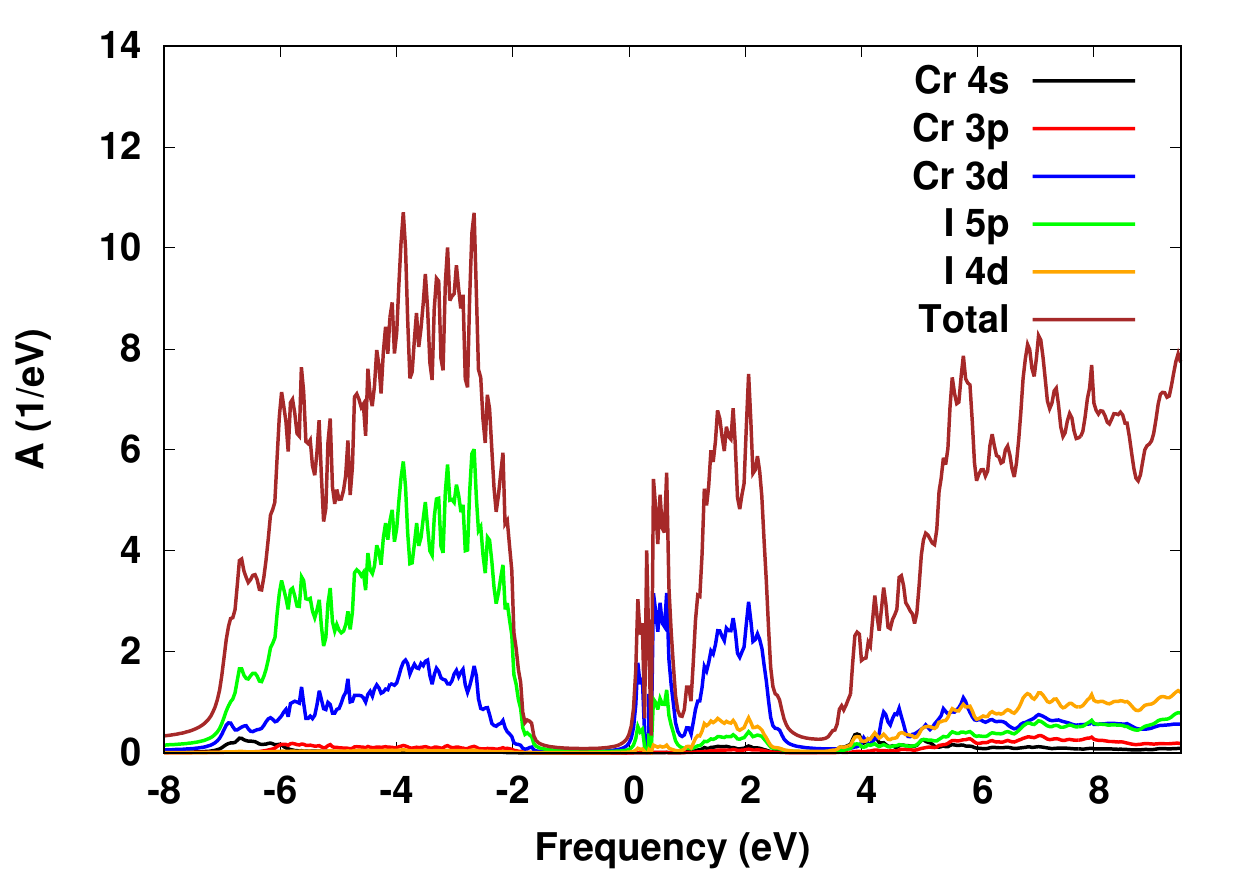}
    \caption{Partial and total spectral functions of CrI$_{3}$ obtained in sc(GW+G3W2) calculation assuming local approximation for the vertex part. Scalar relativistic results. Sums of spin-up and spin-down quantities are given. Analytical continuation of self energy\cite{jltp_29_179,cpc_257_107502} was used to get Green's function on the real frequency axis.}
    \label{psiloc}
\end{figure}

In the evaluation of the band gap, the issues with the local approximation become hidden to a certain degree, as we integrate over the Brillouin zone a few times during every self-consistency iteration. Still, the band gap evaluated in the single site approximation (1.87 eV), tells us that the corresponding correction to the GW value  is almost 25\% larger than the correction obtained without using the local approximation (where the gap is 2.25 eV). The effect of the vertex correction is smaller in the full case because of the interference effects which are neglected in the local approximation. If we forget for a moment about the issues with polarizability and self energy detailed above, the final band gap obtained in the single site approximation might seem reasonable. Partial and total spectral functions obtained with local approximation and shown in Fig. \ref{psiloc} show some differences with the corresponding spectral functions obtained without using the local approximation (Fig. \ref{pdos_sr}, upper right window) but those differences are not dramatic. However, considering the problems with this approximation at the intermediate steps of the calculation, one can conclude that the local approximation (even for the diagrams beyond GW level) represents a poor alternative to the methods which treat the non-local effects systematically. Whereas the quantitative effects are, most likely, material dependent, there is no reason to think that this conclusion will be different for the majority of materials. Considering the importance of this conclusion for GW+DMFT (and related) method, more studies of this kind are needed. As a remedy for the most problematic situations, where both the non-locality effects beyond GW and the strong correlations beyond sc(GW+G3W2) are important, one can suggest an extension of GW+DMFT, for instance sc(GW+G3W2)+DMFT method, which, at least formally, can be implemented along the same lines as GW+DMFT. In this method, DMFT would only be used for evaluation of the diagrams not included in sc(GW+G3W2) approach.

\section*{Conclusions}
\label{concl}

In conclusion, we have applied two self consistent diagrammatic approaches beyond GW approximation to study the electronic structure of the layered van der Waals ferromagnet CrI$_{3}$. Considerable overestimation of the band gap obtained in other works when using G0W0 approach was shown to be remedied by applying the vertex corrections. The important correction comes from the first order vertex function used in both polarizability and self energy. Application of Bethe-Salpeter equation for polarizability further improves the band gap. Inclusion of SOC is important, but its effect is smaller than the effect of vertex corrections.

We also studied the non-locality effects in the diagrams beyond GW approximation and found them as sufficiently large. This can have an impact on development of the methods like GW+DMFT.

As an interesting venue for future work on the subject one can consider studying optical properties of CrI$_{3}$ and other materials using vertex-corrected GW calculations as a starting point for a standard implementation of BSE. "Standard implementation" here means using static (taken at zero frequency) screened interaction W in the kernel of BSE. In standard implementation, one casts BSE in an effective eigen value problem from which the exciton spectra can be directly obtained. Recently, it was shown how it can be done in the context of self-consistent QSGW calculations.\cite{arx_2106_09137}

\section*{Acknowledgments}
\label{acknow}
This work was   supported by the U.S. Department of energy, Office of Science, Basic
Energy Sciences as a part of the Computational Materials Science Program.

\appendix

\section{Details of the Bethe-Salpeter Equation implementation}\label{bse_det}

As it was mentioned in section \ref{meth}, our implementation of BSE uses full frequency dependence of screened interaction W opposite to a common approximation\cite{prl_80_4510,prb_78_085103} where one uses static (frequency independent and taken at zero frequency) screened interaction W. As a result, BSE is solved iteratively in this study. Each iteration adds one more diagram from an infinite sequence shown in Fig. \ref{bse} into the vertex correction to polarizability $\Delta P$. In this Appendix we give the steps of iterations with some details on how frequency/time dependence is handled. Full (and rather lengthy) account of the implementation was published in Ref. [\onlinecite{prb_94_155101}] which includes the details of the basis sets, \textbf{k}-dependencies, and handling of time-to-frequency and frequency-to-time transformations. In this brief account, space arguments of all functions are represented by digits. Integration over repeated space arguments (if they are only on the right hand side of equations) is assumed. Below we use auxiliary functions $K_{0}$, $K$, $\triangle K$ and $\triangle \Gamma$ which are defined by the corresponding equations. Before the iterations we evaluate $K_{0}$:

\begin{align}\label{bse_1}
K_{0}(123;\omega,\nu)=-G(13;\omega)G(32;\omega-\nu),
\end{align}
and assign $\triangle K=0$. $\omega$ and $\nu$ are fermionic and bosonic Matsubara's frequencies correspondingly. Also we transform $K_{0}(123;\tau,\nu)=\frac{1}{\beta}\sum_{\omega}e^{-i\omega\tau}K_{0}(123;\omega,\nu)$ where $\tau$ is Matsubara's time and $\beta=1/T$.

During each iteration we perform the following steps (\ref{bse_2}-\ref{bse_6}):

\begin{align}\label{bse_2}
K(123;\tau,\nu)=K_{0}(123;\tau,\nu)+\triangle K(123;\tau,\nu),
\end{align}

\begin{align}\label{bse_3}
\triangle\Gamma(123;\tau,\nu)=W(21;\tau)K(123;\tau,\nu),
\end{align}

\begin{align}\label{bse_4}
\triangle \Gamma(123;\omega,\nu)=\int d\tau e^{i\omega\tau} \triangle \Gamma(123;\tau,\nu),
\end{align}

\begin{align}\label{bse_5}
\triangle &K(123;\omega,\nu)=\nonumber\\&-G(14;\omega)\triangle\Gamma(453;\omega,\nu) G(52;\omega-\nu),
\end{align}

\begin{align}\label{bse_6}
\triangle K(123;\tau,\nu)=\frac{1}{\beta}\sum_{\omega}e^{-i\omega\tau}\triangle K(123;\omega,\nu).
\end{align}

The above steps are repeated a specific number of times (iterations). In the end of iterations we evaluate vertex correction to polarizability:

\begin{align}\label{bse_7}
\triangle P(12;\nu)=-\triangle K(112;\tau=0,\nu).
\end{align}

For weakly correlated semicondictors the iterations (\ref{bse_2}-\ref{bse_6}) converge very fast (see for instance Fig. 7 in Ref. [\onlinecite{prb_95_195120}]). In case of CrI$_{3}$ we also found that 4 iterations were quite sufficient.


\begin{thebibliography}{47}
\expandafter\ifx\csname natexlab\endcsname\relax\def\natexlab#1{#1}\fi
\expandafter\ifx\csname bibnamefont\endcsname\relax
  \def\bibnamefont#1{#1}\fi
\expandafter\ifx\csname bibfnamefont\endcsname\relax
  \def\bibfnamefont#1{#1}\fi
\expandafter\ifx\csname citenamefont\endcsname\relax
  \def\citenamefont#1{#1}\fi
\expandafter\ifx\csname url\endcsname\relax
  \def\url#1{\texttt{#1}}\fi
\expandafter\ifx\csname urlprefix\endcsname\relax\def\urlprefix{URL }\fi
\providecommand{\bibinfo}[2]{#2}
\providecommand{\eprint}[2][]{\url{#2}}

\bibitem[{\citenamefont{{B.~Huang, G.~Clark, E.~Navarro-Moratalla, D.~R.~Klein,
  R.~Cheng, K.~L.~Seyler, D.~Zhong, E.~Schmidgall, M.~A.~McGuire, D.~H.~Cobden,
  W.~Yao, D.~Xiao, P.~Jarillo-Herrero and X.~Xu}}(2017)}]{nature_546_270}
\bibinfo{author}{\bibnamefont{{B.~Huang, G.~Clark, E.~Navarro-Moratalla,
  D.~R.~Klein, R.~Cheng, K.~L.~Seyler, D.~Zhong, E.~Schmidgall, M.~A.~McGuire,
  D.~H.~Cobden, W.~Yao, D.~Xiao, P.~Jarillo-Herrero and X.~Xu}}},
  \bibinfo{journal}{Nature} \textbf{\bibinfo{volume}{546}},
  \bibinfo{pages}{270} (\bibinfo{year}{2017}).

\bibitem[{\citenamefont{{Y.~Liu, L.~Wu, X.~Tong, J.~Li, J.~Tao, Y.~Zhu and
  C.~Petrovic}}(2019)}]{scirep_9_13599}
\bibinfo{author}{\bibnamefont{{Y.~Liu, L.~Wu, X.~Tong, J.~Li, J.~Tao, Y.~Zhu
  and C.~Petrovic}}}, \bibinfo{journal}{Sci.~Rep.}
  \textbf{\bibinfo{volume}{9}}, \bibinfo{pages}{13599} (\bibinfo{year}{2019}).

\bibitem[{\citenamefont{{A.~K.~Kundu, Y.~Liu, C.~Petrovic and
  T.~Valla}}(2020)}]{scirep_10_15602}
\bibinfo{author}{\bibnamefont{{A.~K.~Kundu, Y.~Liu, C.~Petrovic and
  T.~Valla}}}, \bibinfo{journal}{Sci.~Rep.} \textbf{\bibinfo{volume}{10}},
  \bibinfo{pages}{15602} (\bibinfo{year}{2020}).

\bibitem[{\citenamefont{{J.~F.~Dillon, Jr. and
  C.~E.~Olson}}(1965)}]{jap_36_1259}
\bibinfo{author}{\bibnamefont{{J.~F.~Dillon, Jr. and C.~E.~Olson}}},
  \bibinfo{journal}{J.~of~Appl.~Phys.} \textbf{\bibinfo{volume}{36}},
  \bibinfo{pages}{1259} (\bibinfo{year}{1965}).

\bibitem[{\citenamefont{{S.~W.~Jang, M.~Y.~Jeong, H.~Yoon, S.~Ryee, and
  M.~J.~Han}}(2019)}]{prm_3_031001}
\bibinfo{author}{\bibnamefont{{S.~W.~Jang, M.~Y.~Jeong, H.~Yoon, S.~Ryee, and
  M.~J.~Han}}}, \bibinfo{journal}{Phys.~Rev.~Materials}
  \textbf{\bibinfo{volume}{3}}, \bibinfo{pages}{031001(R)}
  (\bibinfo{year}{2019}).

\bibitem[{\citenamefont{{W.~-B.~Zhang, Q.~Qu, P.~Zhua and
  C.~-H.~Lam}}(2015)}]{jmcc_3_12457}
\bibinfo{author}{\bibnamefont{{W.~-B.~Zhang, Q.~Qu, P.~Zhua and C.~-H.~Lam}}},
  \bibinfo{journal}{J.~of~Mater.~Chem.~C} \textbf{\bibinfo{volume}{3}},
  \bibinfo{pages}{12457} (\bibinfo{year}{2015}).

\bibitem[{\citenamefont{{V.~K.~Gudelli1 and G.~-Y.~Guo}}(2019)}]{njp_21_053012}
\bibinfo{author}{\bibnamefont{{V.~K.~Gudelli1 and G.~-Y.~Guo}}},
  \bibinfo{journal}{New~J.~of~Phys.} \textbf{\bibinfo{volume}{21}},
  \bibinfo{pages}{053012} (\bibinfo{year}{2019}).

\bibitem[{\citenamefont{{P.~Jiang, L.~Li, Z.~Liao, Y.~X.~Zhao, and
  Z.~Zhong}}(2018)}]{nanolett_18_3844}
\bibinfo{author}{\bibnamefont{{P.~Jiang, L.~Li, Z.~Liao, Y.~X.~Zhao, and
  Z.~Zhong}}}, \bibinfo{journal}{New~J.~of~Phys.}
  \textbf{\bibinfo{volume}{18}}, \bibinfo{pages}{3844} (\bibinfo{year}{2018}).

\bibitem[{\citenamefont{{M.~Wu, Z.~Li, T.~Cao, and
  S.~G.~Louie}}(2019)}]{ncomm_10_2371}
\bibinfo{author}{\bibnamefont{{M.~Wu, Z.~Li, T.~Cao, and S.~G.~Louie}}},
  \bibinfo{journal}{Nature~Comm.} \textbf{\bibinfo{volume}{10}},
  \bibinfo{pages}{2371} (\bibinfo{year}{2019}).

\bibitem[{\citenamefont{{Y.~Lee, T.~Kotani, and L.~Ke}}(2020)}]{prb_101_241409}
\bibinfo{author}{\bibnamefont{{Y.~Lee, T.~Kotani, and L.~Ke}}},
  \bibinfo{journal}{Phys.~Rev.~B} \textbf{\bibinfo{volume}{101}},
  \bibinfo{pages}{241409(R)} (\bibinfo{year}{2020}).

\bibitem[{\citenamefont{{H.~Jiang, and P.~Blaha}}(2016)}]{prb_93_115203}
\bibinfo{author}{\bibnamefont{{H.~Jiang, and P.~Blaha}}},
  \bibinfo{journal}{Phys.~Rev.~B} \textbf{\bibinfo{volume}{93}},
  \bibinfo{pages}{115203} (\bibinfo{year}{2016}).

\bibitem[{\citenamefont{{A.~Molina-Sanchez, G.~Catarina, D.~Sangallib and
  J.~Fernandez-Rossier}}(2020)}]{jmcc_8_8856}
\bibinfo{author}{\bibnamefont{{A.~Molina-Sanchez, G.~Catarina, D.~Sangallib and
  J.~Fernandez-Rossier}}}, \bibinfo{journal}{J.Mater.~Chem.~C}
  \textbf{\bibinfo{volume}{8}}, \bibinfo{pages}{8856} (\bibinfo{year}{2020}).

\bibitem[{\citenamefont{{D.~Deguchi, K.~Sato, H.~Kino, and
  T.~Kotani}}(2016)}]{jjap_55_051201}
\bibinfo{author}{\bibnamefont{{D.~Deguchi, K.~Sato, H.~Kino, and T.~Kotani}}},
  \bibinfo{journal}{Japan.~J.~of~Appl.~Phys.} \textbf{\bibinfo{volume}{55}},
  \bibinfo{pages}{051201} (\bibinfo{year}{2016}).

\bibitem[{\citenamefont{{C.~Bhandari, M.~van~Schilfgaarde, T.~Kotani,
  W.~R.~L.~Lambrecht}}(2018)}]{prm_2_013807}
\bibinfo{author}{\bibnamefont{{C.~Bhandari, M.~van~Schilfgaarde, T.~Kotani,
  W.~R.~L.~Lambrecht}}}, \bibinfo{journal}{Phys.~Rev.~Materials}
  \textbf{\bibinfo{volume}{2}}, \bibinfo{pages}{013807} (\bibinfo{year}{2018}).

\bibitem[{\citenamefont{{S.~Biermann, F.~Aryasetiawan, and
  A.~Georges}}(2003)}]{prl_90_086402}
\bibinfo{author}{\bibnamefont{{S.~Biermann, F.~Aryasetiawan, and A.~Georges}}},
  \bibinfo{journal}{Phys.~Rev.~Lett.} \textbf{\bibinfo{volume}{90}},
  \bibinfo{pages}{086402} (\bibinfo{year}{2003}).

\bibitem[{\citenamefont{{L.~Boehnke, F.~Nilsson, F.~Aryasetiawan, and
  P.~Werner}}(2016)}]{prb_94_201106}
\bibinfo{author}{\bibnamefont{{L.~Boehnke, F.~Nilsson, F.~Aryasetiawan, and
  P.~Werner}}}, \bibinfo{journal}{Phys.~Rev.~B} \textbf{\bibinfo{volume}{94}},
  \bibinfo{pages}{201106(R)} (\bibinfo{year}{2016}).

\bibitem[{\citenamefont{{F.~Nilsson, L.~Boehnke, P.~Werner, and
  F.~Aryasetiawan}}(2017)}]{prm_1_043803}
\bibinfo{author}{\bibnamefont{{F.~Nilsson, L.~Boehnke, P.~Werner, and
  F.~Aryasetiawan}}}, \bibinfo{journal}{Phys.~Rev.~Materials}
  \textbf{\bibinfo{volume}{1}}, \bibinfo{pages}{043803} (\bibinfo{year}{2017}).

\bibitem[{\citenamefont{{F.~Petocchi, F.~Nilsson, F.~Aryasetiawan, and
  P.~Werner}}(2020)}]{prr_2_013191}
\bibinfo{author}{\bibnamefont{{F.~Petocchi, F.~Nilsson, F.~Aryasetiawan, and
  P.~Werner}}}, \bibinfo{journal}{Phys.~Rev.~Research}
  \textbf{\bibinfo{volume}{2}}, \bibinfo{pages}{013191} (\bibinfo{year}{2020}).

\bibitem[{\citenamefont{{F.~Petocchi, V.~Christiansson, F.~Nilsson,
  F.~Aryasetiawan, and P.~Werner}}(2020)}]{prx_10_041047}
\bibinfo{author}{\bibnamefont{{F.~Petocchi, V.~Christiansson, F.~Nilsson,
  F.~Aryasetiawan, and P.~Werner}}}, \bibinfo{journal}{Phys.~Rev.~X}
  \textbf{\bibinfo{volume}{10}}, \bibinfo{pages}{041047}
  (\bibinfo{year}{2020}).

\bibitem[{\citenamefont{{S.~Choi, A.~Kutepov, K.~Haule, M.~van~Schilfgaarde,
  and G.~Kotliar}}(2016)}]{npj_1_16001}
\bibinfo{author}{\bibnamefont{{S.~Choi, A.~Kutepov, K.~Haule,
  M.~van~Schilfgaarde, and G.~Kotliar}}}, \bibinfo{journal}{NPJ Quantum
  Materials} \textbf{\bibinfo{volume}{1}}, \bibinfo{pages}{16001}
  (\bibinfo{year}{2016}).

\bibitem[{\citenamefont{{S.~Choi, P.~Semon, B.~Kang, A.~Kutepov,
  G.~Kotliar}}(2019)}]{cpc_244_277}
\bibinfo{author}{\bibnamefont{{S.~Choi, P.~Semon, B.~Kang, A.~Kutepov,
  G.~Kotliar}}}, \bibinfo{journal}{Comp.~Phys.~Comm.}
  \textbf{\bibinfo{volume}{244}}, \bibinfo{pages}{277} (\bibinfo{year}{2019}).

\bibitem[{fla()}]{flapwmbpt_1}
\bibinfo{note}{The latest publicly available version of the FlapwMBPT code
  (FlapwMBPT2106) can be downloaded from the website
  https://github.com/andreykutepov65/FlapwMBPT.}

\bibitem[{\citenamefont{{A.~L.~Kutepov}}(2021{\natexlab{a}})}]{prb_103_165101}
\bibinfo{author}{\bibnamefont{{A.~L.~Kutepov}}},
  \bibinfo{journal}{Phys.~Rev.~B} \textbf{\bibinfo{volume}{103}},
  \bibinfo{pages}{165101} (\bibinfo{year}{2021}{\natexlab{a}}).

\bibitem[{\citenamefont{{A.~L.~Kutepov}}(2021{\natexlab{b}})}]{jcm_33_235503}
\bibinfo{author}{\bibnamefont{{A.~L.~Kutepov}}}, \bibinfo{journal}{J.~Phys.:
  Condens.~Matter} \textbf{\bibinfo{volume}{33}}, \bibinfo{pages}{235503}
  (\bibinfo{year}{2021}{\natexlab{b}}).

\bibitem[{\citenamefont{{J.~P.~Perdew and Y.~Wang}}(1992)}]{prb_45_13244}
\bibinfo{author}{\bibnamefont{{J.~P.~Perdew and Y.~Wang}}},
  \bibinfo{journal}{Phys.~Rev.B} \textbf{\bibinfo{volume}{45}},
  \bibinfo{pages}{13244} (\bibinfo{year}{1992}).

\bibitem[{\citenamefont{{L.~Hedin}}(1965)}]{pr_139_A796}
\bibinfo{author}{\bibnamefont{{L.~Hedin}}}, \bibinfo{journal}{Phys.~Rev.}
  \textbf{\bibinfo{volume}{139}}, \bibinfo{pages}{A796} (\bibinfo{year}{1965}).

\bibitem[{\citenamefont{{A.~L.~Kutepov}}(2021{\natexlab{c}})}]{arx_2105_03770}
\bibinfo{author}{\bibnamefont{{A.~L.~Kutepov}}},
  \bibinfo{journal}{arXiv.cond.mat.:2105.03770}
  (\bibinfo{year}{2021}{\natexlab{c}}).

\bibitem[{\citenamefont{{C.-O.~Almbladh, U.~von~Barth and
  R.~van~Leeuwen}}(1999)}]{ijmpb_13_535}
\bibinfo{author}{\bibnamefont{{C.-O.~Almbladh, U.~von~Barth and
  R.~van~Leeuwen}}}, \bibinfo{journal}{Int.~J. of Mod.Phys. B}
  \textbf{\bibinfo{volume}{13}}, \bibinfo{pages}{535} (\bibinfo{year}{1999}).

\bibitem[{\citenamefont{{A.~L.~Kutepov}}(2016)}]{prb_94_155101}
\bibinfo{author}{\bibnamefont{{A.~L.~Kutepov}}},
  \bibinfo{journal}{Phys.~Rev.~B} \textbf{\bibinfo{volume}{94}},
  \bibinfo{pages}{155101} (\bibinfo{year}{2016}).

\bibitem[{\citenamefont{{S.~Albrecht, L.~Reining, R.~Del~Sole and
  G.~Onida}}(1998)}]{prl_80_4510}
\bibinfo{author}{\bibnamefont{{S.~Albrecht, L.~Reining, R.~Del~Sole and
  G.~Onida}}}, \bibinfo{journal}{Phys. Rev. Lett.}
  \textbf{\bibinfo{volume}{80}}, \bibinfo{pages}{4510} (\bibinfo{year}{1998}).

\bibitem[{\citenamefont{{F.~Fuchs, C.~R\"{o}dl, A.~Schleife, and
  F.~Bechstedt}}(2008)}]{prb_78_085103}
\bibinfo{author}{\bibnamefont{{F.~Fuchs, C.~R\"{o}dl, A.~Schleife, and
  F.~Bechstedt}}}, \bibinfo{journal}{Phys.~Rev.~B}
  \textbf{\bibinfo{volume}{78}}, \bibinfo{pages}{085103}
  (\bibinfo{year}{2008}).

\bibitem[{\citenamefont{{A.~L.~Kutepov}}(2017)}]{prb_95_195120}
\bibinfo{author}{\bibnamefont{{A.~L.~Kutepov}}},
  \bibinfo{journal}{Phys.~Rev.~B} \textbf{\bibinfo{volume}{95}},
  \bibinfo{pages}{195120} (\bibinfo{year}{2017}).

\bibitem[{\citenamefont{{A.~L.~Kutepov and G.~Kotliar}}(2017)}]{prb_96_035108}
\bibinfo{author}{\bibnamefont{{A.~L.~Kutepov and G.~Kotliar}}},
  \bibinfo{journal}{Phys.~Rev.~B} \textbf{\bibinfo{volume}{96}},
  \bibinfo{pages}{035108} (\bibinfo{year}{2017}).

\bibitem[{\citenamefont{{A.~Kutepov, K.~Haule, S.~Y.~Savrasov, and
  G.~Kotliar}}(2012)}]{prb_85_155129}
\bibinfo{author}{\bibnamefont{{A.~Kutepov, K.~Haule, S.~Y.~Savrasov, and
  G.~Kotliar}}}, \bibinfo{journal}{Phys.~Rev.~B} \textbf{\bibinfo{volume}{85}},
  \bibinfo{pages}{155129} (\bibinfo{year}{2012}).

\bibitem[{\citenamefont{{A.~L.~Kutepov, V.~S.~Oudovenko,
  G.~Kotliar}}(2017)}]{cpc_219_407}
\bibinfo{author}{\bibnamefont{{A.~L.~Kutepov, V.~S.~Oudovenko, G.~Kotliar}}},
  \bibinfo{journal}{Comp.~Phys.~Comm.} \textbf{\bibinfo{volume}{219}},
  \bibinfo{pages}{407} (\bibinfo{year}{2017}).

\bibitem[{\citenamefont{{T.~Kotani and M.~van~Schilfgaarde,
  S.~V.~Faleev}}(2007)}]{prb_76_165106}
\bibinfo{author}{\bibnamefont{{T.~Kotani and M.~van~Schilfgaarde,
  S.~V.~Faleev}}}, \bibinfo{journal}{Phys.~Rev.B}
  \textbf{\bibinfo{volume}{76}}, \bibinfo{pages}{165106}
  (\bibinfo{year}{2007}).

\bibitem[{\citenamefont{{T.~Takeda}}(1978)}]{zpb_32_43}
\bibinfo{author}{\bibnamefont{{T.~Takeda}}}, \bibinfo{journal}{Z.~Physik~B}
  \textbf{\bibinfo{volume}{32}}, \bibinfo{pages}{43} (\bibinfo{year}{1978}).

\bibitem[{\citenamefont{{M.~A.~McGuire, H.~Dixit, V.~R.~Cooper, and
  B.~C.~Sales}}(2015)}]{cm_27_612}
\bibinfo{author}{\bibnamefont{{M.~A.~McGuire, H.~Dixit, V.~R.~Cooper, and
  B.~C.~Sales}}}, \bibinfo{journal}{Chem.~Mater.}
  \textbf{\bibinfo{volume}{27}}, \bibinfo{pages}{612} (\bibinfo{year}{2015}).

\bibitem[{\citenamefont{{J.P.~Perdew, K.~Burke and
  M.~Ernzerhof}}(1996)}]{prl_77_3865}
\bibinfo{author}{\bibnamefont{{J.P.~Perdew, K.~Burke and M.~Ernzerhof}}},
  \bibinfo{journal}{Phys. Rev. Lett.} \textbf{\bibinfo{volume}{77}},
  \bibinfo{pages}{3865} (\bibinfo{year}{1996}).

\bibitem[{\citenamefont{{M.~Grumet, P.~Liu, M.~Kaltak, J.~Klimes, and
  G.~Kresse}}(2018)}]{prb_98_155143}
\bibinfo{author}{\bibnamefont{{M.~Grumet, P.~Liu, M.~Kaltak, J.~Klimes, and
  G.~Kresse}}}, \bibinfo{journal}{Phys.~Rev.~B} \textbf{\bibinfo{volume}{98}},
  \bibinfo{pages}{155143} (\bibinfo{year}{2018}).

\bibitem[{\citenamefont{{H.~J.~Vidberg and J.~W.~Serene}}(1977)}]{jltp_29_179}
\bibinfo{author}{\bibnamefont{{H.~J.~Vidberg and J.~W.~Serene}}},
  \bibinfo{journal}{J.~Low~Temp.~Phys.} \textbf{\bibinfo{volume}{29}},
  \bibinfo{pages}{179} (\bibinfo{year}{1977}).

\bibitem[{\citenamefont{{A.~L.~Kutepov}}(2020)}]{cpc_257_107502}
\bibinfo{author}{\bibnamefont{{A.~L.~Kutepov}}},
  \bibinfo{journal}{Comp.~Phys.~Commun.} \textbf{\bibinfo{volume}{257}},
  \bibinfo{pages}{107502} (\bibinfo{year}{2020}).

\bibitem[{\citenamefont{{G.~Stefanucci, Y.~Pavlyukh, A.~-M.~Uimonen,
  R.~van~Leeuwen}}(2014)}]{prb_90_115134}
\bibinfo{author}{\bibnamefont{{G.~Stefanucci, Y.~Pavlyukh, A.~-M.~Uimonen,
  R.~van~Leeuwen}}}, \bibinfo{journal}{Phys.~Rev.~B}
  \textbf{\bibinfo{volume}{90}}, \bibinfo{pages}{115134}
  (\bibinfo{year}{2014}).

\bibitem[{\citenamefont{{A.~-M.~Uimonen, G.~Stefanucci, Y.~Pavlyukh, and
  R.~van~Leeuwen}}(2015)}]{prb_91_115104}
\bibinfo{author}{\bibnamefont{{A.~-M.~Uimonen, G.~Stefanucci, Y.~Pavlyukh, and
  R.~van~Leeuwen}}}, \bibinfo{journal}{Phys.~Rev.~B}
  \textbf{\bibinfo{volume}{91}}, \bibinfo{pages}{115104}
  (\bibinfo{year}{2015}).

\bibitem[{\citenamefont{{Y.~Pavlyukh, G.~Stefanucci,
  R.~van~Leeuwen}}(2020)}]{prb_102_045121}
\bibinfo{author}{\bibnamefont{{Y.~Pavlyukh, G.~Stefanucci, R.~van~Leeuwen}}},
  \bibinfo{journal}{Phys.~Rev.~B} \textbf{\bibinfo{volume}{102}},
  \bibinfo{pages}{045121} (\bibinfo{year}{2020}).

\bibitem[{\citenamefont{{A.~L.~Kutepov}}(2021{\natexlab{d}})}]{arx_2106_03800}
\bibinfo{author}{\bibnamefont{{A.~L.~Kutepov}}},
  \bibinfo{journal}{arXiv.cond.mat.:2106.03800}
  (\bibinfo{year}{2021}{\natexlab{d}}).

\bibitem[{\citenamefont{{S.~K.~Radha, W.~R.~L.~Lambrecht, B.~Cunningham,
  M.~Gr\"{u}ning, D.~Pashov and M.~van~Schilfgaarde}}(2021)}]{arx_2106_09137}
\bibinfo{author}{\bibnamefont{{S.~K.~Radha, W.~R.~L.~Lambrecht, B.~Cunningham,
  M.~Gr\"{u}ning, D.~Pashov and M.~van~Schilfgaarde}}},
  \bibinfo{journal}{arXiv.cond.mat.:2106.09137}  (\bibinfo{year}{2021}).

\end{thebibliography}

\end{document}